\documentclass[prl,twocolumn,groupedaddress,footinbib]{revtex4-1}

\usepackage{amsmath}
\usepackage{hyperref}
\usepackage{nicefrac}
\usepackage{cleveref}
\usepackage{xcolor}

\input{scrload}

\usepackage{bm}
\usepackage{soul}

\newcommand{\br}{{\bm r}}
\newcommand{\bu}{{\bm u}}
\newcommand{\bj}{{\bm j}}
\newcommand{\bq}{{\bm q}}

\newcommand{\ud}{{\mathrm d}} 

\usepackage{changebar}

\begin{document}
\title{Thermal forces from a microscopic perspective}

\author{Pietro Anzini} 
\author{Gaia Maria Colombo}
\author{Zeno Filiberti} 
\author{Alberto Parola}
\email[Corresponding author ]{alberto.parola@uninsubria.it}
\affiliation{Dipartimento di Scienza e Alta Tecnologia, Universit\`a degli Studi dell'Insubria, Como, Italy}

\begin{abstract} 
Thermal gradients lead to macroscopic fluid motion if a confining 
surface is present along the gradient. This fundamental nonequilibrium effect, known as \hbox{thermo-osmosis}, is held responsible for 
particle thermophoresis in colloidal suspensions. A unified approach for \hbox{thermo-osmosis} in liquids and in gases is still lacking.
Linear Response Theory is generalised to inhomogeneous systems, leading to an exact microscopic theory for the \hbox{thermo-osmotic} flow
showing that the
effect originates from two independent physical mechanisms, playing different roles in the gas and liquid phases, reducing
to known expressions in the appropriate limits.
\end{abstract}

\maketitle
When a uniform bulk fluid is placed in a thermal gradient mechanical equilibrium quickly sets in via the
force balance condition, implying constant pressure throughout the system. In the absence of external forces,
the \hbox{steady-state} is characterised by a space dependent density profile and a constant heat flux not associated 
to mass current~\cite{*[{See e.g. }] [{, Sec. 4.}] landau_fluid}.
The action of a thermal gradient on a fluid then resembles the effect of a fictitious ``thermal force''~\cite{luttinger_64},
which has been known to play also a {\it dynamic} role since 
the first studies in gases by Feddersen, Crookes, Reynolds and Maxwell \footnote{See Ref.s~\cite{feddersen_1972,crookes_radiometer,reynolds_1879,maxwell_1879}.
An historical account on the controversies related to the physical understanding of the behaviour of the radiometer
can be found in Ref.s~\cite{maxwell_reynolds_radiometer,piazza_2004}.}. 
The onset of a stationary fluid flow induced by temperature gradients (in the absence of symmetry breaking forces like gravity, when convection
dominates) is named \hbox{{\it thermo-osmosis}} \footnote{Up to our knowledge, the term \hbox{{\it thermo-osmosis}} (or thermal osmosis)
has been coined by Lippman~\cite{lippmann_1907} in the case of liquids. 
In gases the same phenomenon is referred to as {\it thermal creep} or {\it thermal transpiration} (after Reynolds).} 
and only occurs due to the presence of a confining 
surface parallel to the thermal gradient, as already pointed out both in gases~\cite{feddersen_1972,reynolds_1879,maxwell_1879} 
and in liquids~\cite{lippmann_1907,aubert_1912,derjaguin_sidorenkov_1941thermoosmosis}.
\hbox{Thermo-osmosis} is believed to be the driving mechanism for thermophoresis, i.e. the motion of a colloidal particle in 
a solvent due to a temperature gradient~\cite{rev_thermo_gas,pp,wurger_2010}, where the slip of the fluid in the boundary layer close to the 
particle's surface gives rise to momentum transfer and eventually to particle motion. 
\hbox{Thermo-osmosis} is therefore one of the most fundamental manifestations of thermal forces and its physical origin is deeply rooted 
in nonequilibrium statistical mechanics. At the same time it is of great interest for applications as a mechanism for 
governing particle's motion at the nanoscale~\cite{swimmers_2007,mitocondri}.

A unified description of \hbox{thermo-osmosis} 
is still lacking: The phenomenon was theoretically investigated mainly in the gas phase, 
where the fluid moves from the cold to the hot side and the characteristic lengthscale   
is of the order of the molecular mean free path~\cite{sone_2000}.
The kinetic theory of gases has been used in this framework
since the seminal work by Maxwell~\cite{maxwell_1879}, who showed 
that the thermal creep is due to the tangential stress exerted by the gas on the fixed confining surface
in the direction opposite to the temperature gradient. 
Such a stress, however, requires some exchange of energy and tangential momentum in the \hbox{wall-particle} 
scattering process and therefore 
depends on the 
modelling of \hbox{fluid-surface} interactions. 
\hbox{Thermo-osmosis} in the liquid regime is 
considerably less studied, both theoretically~\cite{eastman26,derjaguin_sidorenkov_1941thermoosmosis,degroot47,denbigh1952} and 
experimentally~\cite{derjaguin_sidorenkov_1941thermoosmosis,hutchisonnixondenbigh_1948,gaeta,bregulla_2016}.
In addition, as shown in the recent review \cite{barragankjelstrup_2016review}, experiments often disagree 
even about the direction of the \hbox{thermo-osmotic} flow.
Nonequilibrium irreversible thermodynamics, based on the concept of local thermal equilibrium, was first used by Derjaguin 
to relate the \hbox{thermo-osmotic} velocity in liquids to the change of the local enthalpy of the fluid near the 
confining surface~\cite{derjaguin_sidorenkov_1941thermoosmosis,surface_forces_1987}. 
Then, according to this macroscopic approach the physical origin of the fluid motion is 
due to the modification in the local thermodynamic properties of the fluid induced 
by the presence of a wall, as pointed out in Ref.s~\cite{anderson_review,han}. 
Clearly, in the rarefied limit, which Derjaguin does not consider, the argument 
must fail because the effects of a hard wall on the (local) equilibrium properties of the gas disappear 
at low density. Only recently numerical simulations directly tackled this subtle nonequilibrium problem in the 
liquid regime~\cite{bjorn1999,galliero2002,ganti17,Ganti18,joly17,joly18}, but a clear numerical evidence 
of the correctness of the Derjaguin formula has not been established yet. 
Rather, in Ref.~\cite{ganti17} it was pointed out that Derjaguin expression cannot be correct because 
neither the enthalpy density nor the tangential pressure close to a surface is well defined on microscopic grounds.

This unsatisfactory 
setting calls for a first principle approach to the phenomenon, able to quantitatively evaluate the extent of the 
\hbox{thermo-osmotic} slip in terms of well defined properties of the fluid which can be measured in experiments 
and calculated in numerical simulations.
In this Letter we present a microscopic description of \hbox{thermo-osmosis} on the basis of 
statistical physics: 
Linear Response Theory generalised to inhomogeneous and anisotropic environments. In the case of
an imposed  uniform thermal gradient, the use of conservation laws allows to evaluate the 
velocity profile of the fluid and the \hbox{thermo-osmotic} slip in terms of 
both the static and the dynamic 
equilibrium properties of the fluid near the surface:
In the appropriate limits the well known expressions obtained within kinetic theory and nonequilibrium thermodynamics (Derjaguin) are recovered
by retaining each of these terms, showing that the gas and liquid regimes are indeed 
governed by different physical mechanisms. 

The \hbox{Green-Kubo} formalism for linear response theory~\cite{kubo,zwanzig} was generalised by Mori to deal with the thermal 
transport coefficients~\cite{mori56,mori58}. The starting point, as in the nonequilibrium thermodynamics
framework, is the concept of Local Equilibrium (LE) mathematically defined by the many body distribution function 
\begin{equation}
F^{LE} = {\cal Q}^{-1} \, \mathrm{e}^{-\int \ud\br \,\beta(\br)\,\hat{\cal E}(\br)},
\label{le}
\end{equation}
where ${\cal Q}$ is the partition function and the local energy density $\hat{\cal E}(\br)$ is expressed in terms of the conserved densities as
\begin{equation}
\hat{\cal E}(\br) = \hat{\cal H}(\br) -\bu(\br)\cdot\hat\bj(\br) -\mu(\br)\hat\rho(\br).
\notag
\end{equation}
Here $\beta(\br), \bu(\br)$ and $\mu(\br)$ are external fields governing the temperature profile,
the fluid velocity and the local chemical potential blue (per unit mass) respectively. $\hat{\cal H}(\br)$ is the 
microscopic many body Hamiltonian density
\begingroup\makeatletter\def\f@size{10}\check@mathfonts 
\medmuskip=1mu
\thinmuskip=2mu
\thickmuskip=1mu
\begin{equation} 
\hat{\cal H}(\br)=
\sum_i \delta(\bq_i-\br)\Bigg[\! \frac{p_i^2}{2m} + \frac{1}{2}\! \sum_{j (\ne i)}\!\! v(|\bq_i-\bq_j|) +V(\bq_i)\!\Bigg]
\label{ham}
\end{equation}
\endgroup
which describes a system of interacting point particles of mass $m$ confined by hard walls represented by the 
external potential $V(\br)$.
The operators  
\begin{eqnarray}
\hat\rho(\br) &=& m\,\sum_i \delta(\bq_i-\br), \notag \\
\hat j^{\alpha}(\br) &=& \sum_i \delta(\bq_i-\br)\,p^{\alpha}_i,
\label{eq:def_momdens}
\end{eqnarray}
define the local mass and momentum densities, which, together with the Hamiltonian density 
$\hat{\cal H}(\br)$ introduced in Eq.~(\ref{ham}), satisfy microscopic conservation equations of the general form
\begin{equation}
\frac{\ud \hat A(\br)}{\ud t}  + \partial_\alpha \hat J_A^{\alpha}(\br) = 0,
\label{eq:def_fluxes}
\end{equation}
where $\hat A(\br)$ is the conserved density whereas $\hat J^\alpha_A(\br)$ is the corresponding current operator.
Here and in the following Greek indices represent spatial components of vectors and tensors and Einstein summation convention
is understood. In our case $\hat J^\alpha_A(\br)$ represents 
the mass current $\hat j^\alpha_{\rho}(\br)$, the momentum 
$\hat J_j^{\alpha\gamma}(\br)$ and the energy flux $\hat J^\alpha_{\cal H}(\br)$ respectively. 
The explicit expressions for the current operators in terms of the coordinates and momenta of the 
particles~\setcounter{footnote}{98}\footnote{\label{foot:definizione_flussi}Note that the conservation equations just define 
the divergence of the current operators. In some case, notably
for the pressure tensor and for the heat flux, this fact poses some mathematical ambiguity in the precise definition of the current as
discussed in detail elsewhere (see Ref.s~\cite{henderson,baroni_15}  and Supplementary Material). In the following we will just assume that the
adopted definition of the current preserves
the short range nature of the dynamic correlations and the spatial symmetries of the system.}
are reported in the Supplementary Material (SM). 

The previously defined local equilibrium distribution function~(\ref{le}) is 
not a solution of the Liouville equation and therefore it cannot describe a stationary state: Even if the
system is initially set in a LE state, its distribution function changes in time in order to reach full thermodynamic equilibrium. 
External constraints may however keep the system out of equilibrium, for instance by enforcing different temperatures at the boundaries,
leading instead to a nonequilibrium stationary state characterised by constant fluxes of particles and/or energy and momentum. 
Accordingly, $F^{LE}$ cannot be used to evaluate 
averages in the resulting stationary state, rather we have to include a correction term coming from the ensuing dynamics. 
It is precisely such a contribution which defines the microscopic expressions of the standard transport coefficients~\cite{mori58,zwanzig,balescu}. 
Within Linear Response Theory the formal expression of the distribution function is known and reads
\begin{eqnarray}
F &=& F^{LE} +  F^{eq}\,\int_0^t \ud t^\prime \int \ud\br\, \beta(\br)\,\Big [ \partial_\alpha\,\hat J^\alpha_{\cal H}(\br,t^\prime)\nonumber \\
&-& u^\alpha(\br)\,\partial_\gamma\, \hat J_j^{\alpha\gamma}(\br,t^\prime) -  
\mu(\br)\,\partial_\alpha\,\hat{j}_{\rho}^\alpha(\br,t^\prime) \Big ],
\label{lrt}
\end{eqnarray}
where $F^{eq}$ is the underlying grand canonical equilibrium distribution function defined by the  
average value of the (inverse) temperature $\beta$ and of the chemical potential (per unit mass) $\mu$  and 
the time dependence of the current operators means that they are evaluated after a time 
lapse $t^\prime$ from the initial configuration.
Averages in the stationary state can be formally evaluated starting from Eq.~(\ref{lrt}),
performing an integration by parts and taking the $t\to\infty$ limit~\footnote{Special care must be paid 
in taking the \hbox{long-time} limit for correlation functions involving currents of
conserved quantities, as in our cases~\cite{simpleliquids_fourth}.}. 
Here we stress that the dynamic corrections in~(\ref{lrt}) only involve the divergence of the fluxes
introduced in~(\ref{eq:def_fluxes}) and the resulting physical averages can be evaluated without ambiguity,
even if the microscopic definition of the current operators is not unique~\cite{Note99}.
For future reference we report the final result for the momentum density $\big< \hat{\bj}(\br)\big>$ to linear 
order in the velocity field $\bu(\br)$ and in the spatial derivatives of the temperature
and the chemical potential:
\begingroup\makeatletter\def\f@size{9}\check@mathfonts
\begin{widetext}
\medmuskip=1mu
\thinmuskip=1mu
\thickmuskip=1mu
\begin{equation}
\big<\hat j^\alpha(\br)\big> =  \big<\hat j^\alpha(\br)\big>_{LE} +
\int_0^\infty \!\!\!\!\! \! \!\!\!\!\!\!\! \ud t \! \int \!\! \ud\br^\prime 
\bigg[ \big< \hat j^\alpha(\br,t) \hat J_{\cal H}^\gamma(\br^\prime)\big>_0 \, \partial_\gamma\beta(\br^\prime) 
-\big< \hat j^\alpha(\br,t) \,\hat j_{\rho}^\gamma(\br^\prime)\big>_0  \partial_\gamma \big[\beta\mu\big](\br^\prime) 
-\big< \hat j^\alpha(\br,t)\, \hat J_j^{\nu\gamma}(\br^\prime)\big>_0 \partial_\gamma \big[\beta u^\nu\big](\br^\prime)\bigg].
\label{momenth}
\end{equation}
\end{widetext}
\endgroup
The averages $\langle\,\dots\rangle_0$ have been evaluated by means of the underlying equilibrium distribution $F^{eq}$, 
and, to linear order in the velocity field, the LE distribution (\ref{le}) gives
$\langle \,\hat j^\alpha(\br) \rangle_{LE} = \rho(\br) \, u^\alpha(\br).$
Equation~(\ref{momenth}) is the formal expression of the \hbox{thermo-osmotic} slip in the presence of a \hbox{non-uniform} temperature field.
Notice that~(\ref{momenth}) also involves {\it \hbox{odd-rank}} tensors, forbidden by space isotropy, 
because this general theoretical framework also applies for a fluid close to an external surface, e.g. a hard wall,
which breaks isotropy defining a preferred direction. 
\newline
Similar formulas can be derived for the averages of 
other physical quantities. 
The LE average of the momentum flux operator $\hat J_j^{\alpha\gamma}(\br)$ gives the pressure tensor at equilibrium 
evaluated at the local temperature and chemical potential, but may also include a \hbox{non-vanishing} \hbox{off-di}agonal contribution, as 
detailed in the SM. 

However, in any experiment the external fields $\beta(\br), \bu(\br)$ and $\mu(\br)$  appearing in Eq.~(\ref{momenth}) cannot be 
fixed from the outset but are rather \hbox{self-consistently} determined by the system,
while the experimental \hbox{set-up} just defines the appropriate 
boundary conditions. Only pressure, temperature and velocity at the boundaries are given, while the spatial variation of the 
same quantities throughout the sample follow from the conservation equations: In \hbox{steady-state} conditions
the divergence of the {\it average} particle, momentum and energy flux must therefore vanish. 
These constraints provide five differential equations for the five external fields appearing in the LE distribution function~(\ref{le}) leading to 
the formal solution of the problem. 
\newline
To proceed further, let us consider a simple ``slab geometry'' 
where the fluid is confined between two infinite hard walls placed at a distance $h$ along the \hbox{$z$-direction}. 
The equilibrium density profile $\rho(\bm r)$ is $z$-dependent and the only \hbox{non-vanishing} components of the
equilibrium pressure tensor define the transverse $p^{xx}(z)=p^{yy}(z)=p_\mathrm{T}(z)$ and the normal pressure 
$p^{zz}(z)=p_\mathrm{N}(z)=p$, which is constant and equals the bulk pressure $p$.
Furthermore, the width $h$ is chosen sufficiently large to guarantee that the fluid in the central region can be considered 
to a good approximation unaffected by the presence of the walls (in practice a few molecular diameters are sufficient).
\newline
A solution to the continuity equations is given by constant values of $\partial_x\beta$ and $\partial_x [\beta\mu]$, while
the velocity field $\bu(z)$ is directed along the \hbox{$x$-axis}. 
Under these assumptions and within this simple geometry the stationary continuity equations for the average mass density 
$\langle \hat{\rho}(\br)\rangle$, the energy density $\langle \hat{\cal H}(\br)\rangle$
and the \hbox{$y$-component} of the average momentum density $\langle\hat{j}^{y}(\br)\rangle$ are identically 
satisfied. Furthermore, the 
conservation law for the normal ($z$) component of the momentum density $\langle \hat{j}^{z}(\br)\rangle$
gives rise to the well known hydrostatic equilibrium condition
\begin{equation}
\partial_\alpha \big< \hat{J}_{j}^{\alpha z}(\br)\big>=\partial_z p_{\mathrm{N}}(z)\Big|_{\beta(x),\mu(x)}=0,
\notag
\end{equation}
where the normal pressure is evaluated at the local temperature and chemical potential.
The only \hbox{non-trivial} continuity equation comes from the conservation of the \hbox{$x$-component} of the momentum density, which 
must be solved imposing that no pressure gradient is present far from the walls {\it (open channel)}. 
The latter condition implies that $\partial_x [\beta\mu]$ can be expressed in terms of the temperature 
gradient by $\partial_x [\beta\mu]=h_m\,\partial_x\beta$, 
where $h_m$ is the enthalpy per unit mass of the fluid in the bulk.
The detailed derivation is discussed in the SM. Here we report the final \hbox{integro-differential} equation for the velocity profile:
\begin{equation}
\int_0^h\ud z^\prime \, {\cal K}(z,z^\prime) \, \partial_{z^\prime} u^x(z^\prime) =
\partial_x\beta \, {\cal S}(z).
\label{equ}
\end{equation}
The kernel ${\cal K}(z,z^\prime)$ is related to the local viscosity of the fluid
\begin{equation}
{\cal K}(z,z^\prime) = \beta\,\int_0^\infty \ud t\int\ud\br_\perp^\prime 
\big <\hat J_j^{xz}(\br,t)\, \hat J_j^{xz}(\br^\prime)\big >_0
\nonumber 
\end{equation}
and the source term ${\cal S}(z)$ can be written as the sum of two distinct contributions ${\cal S}(z)={\cal S}_s(z)+{\cal S}_d(z)$,
depending on the static and dynamic equilibrium correlations respectively: 
\begin{eqnarray}
\label{static}
{\cal S}_s(z) &=& 
\int^z_{h/2} \ud z^\prime \, \left.\frac{\partial p_\mathrm{T}(z^\prime)}{\partial \beta}\right|_p   \nonumber \\
&&\qquad  -\int \ud \br^\prime \, (x-x^\prime)\,\big<\hat J_j^{xz}(\br) \, \hat{\cal P}(\br^\prime) \big>_0, \\
{\cal S}_d(z) &=& \int_0^\infty \ud t\int\ud\br^\prime 
\big <\hat J_j^{xz}(\br,t)\, \hat J_Q^{x}(\br^\prime)\big >_0, \qquad
\label{dynamic}
\end{eqnarray}
where we have introduced the
heat flux operator \hbox{$\hat J_Q^{\alpha}(\br) = \hat J_{\cal H}^\alpha(\br)-h_m\,\hat j_{\rho}^{\alpha}(\br)$}~\cite{simpleliquids_fourth}  
and the operator 
\hbox{$\hat{\cal P}(\br)=h_m\,\hat\rho(\br)-\hat{\cal H}(\br)$}, whose average in a homogeneous system at equilibrium 
reduces to the bulk pressure $p$.
Note that both source terms vanish in the bulk, implying \hbox{$\partial_{z} u^x(z)=0$}. 
In the case of a {\it closed channel}, where a pressure gradient along the \hbox{$x$-direction} is present and the integrated
mass current must vanish, the boundary condition should be modified and the results differ from those reported here. 

The solution of this set of equations provides an expression for the gradient of the velocity field $\partial_z u^x(z)$
independent on the particular definition of the fluxes in~(\ref{eq:def_fluxes}),
because the continuity equations only involve divergences of the fluxes (see~\cite{Note99}).
When the result is substituted into Eq.~(\ref{momenth}) the final formula for the mass current is found:
\begin{eqnarray}
\big<&&\hat{j}^x(z)\big> = \rho(z)\,u^x(z)  \nonumber \\
&&  \qquad + \int_0^\infty \ud t \int \ud\br^\prime \Big [ \big< \hat j^x(\br,t) \,\hat J_Q^x(\br^\prime)\big>_0 \,\partial_x\beta  \nonumber \\
&&\qquad \qquad \quad - \beta\,\big< \hat j^x(\br,t)\, \hat J_j^{xz}(\br^\prime)\big>_0\, \partial_{z^\prime} u^x(z')
\Big ] .
\label{momenth2}
\end{eqnarray}
All the contributions appearing in Eq.~(\ref{momenth2}) vanish for a homogeneous system, 
showing that the physical origin of \hbox{thermo-osmosis} relies 
on the existence of a confining surface~\footnote{The first contribution to the integral in Eq.~(\ref{momenth2}) vanishes in a homogeneous fluid
due to the independence of heat and mass current fluctuations. See e.g. Ref.~\cite{landau_fluid}, Sec. 49 or Ref.~\cite{balescu}, Sec. 12.5.}. 
However, the mass flux is not fully determined by Eq.~(\ref{momenth2}) because the 
velocity field (and not only its derivative) appears in the first term. To resolve this ambiguity we have to 
know the mass flux at a given height $z$. This further requirement is not a limitation of the theory but rather a consequence of
the Galilean invariance (along the \hbox{$x$-direction}) of the equilibrium system which, 
in an experimental \hbox{set-up}, is broken by the presence of friction between the 
fluid and the wall~\cite{friction}. Instead, in the simplified model considered here, the wall is represented by an external confining potential 
(a hard wall) which does not modify the tangential ($x$) component of the particles' momenta. Supplementing this solution 
by a suitable (for instance \hbox{{\it no-slip}}) boundary condition for the mass flux, Eq.~(\ref{momenth2}) allows to evaluate the \hbox{thermo-osmotic} flow in slab geometry:
We first have to solve Eq.~(\ref{equ}) for $u^x(z)$ and then substitute the result into Eq.~(\ref{momenth2}).

The above analysis of a model of simple fluid close to a wall is exact, within 
Linear Response Theory, and  shows that two distinct mechanisms give rise to \hbox{thermo-osmosis}, both related to interface physics: 
The presence of anisotropies in the pressure tensor close to the wall (see Eq.~(\ref{static})) and the effect of a confining surface on 
the dynamic correlation functions (see Eq.~(\ref{dynamic})). 
We now consider two limiting situations where 
these terms play a very different role in order to clarify their relevance in providing the required thermal force. 

In liquids we expect that the correlations can be estimated by their bulk value 
and the kernel ${\cal K}(z,z^\prime)$ is taken to be a short-ranged function
\begin{equation}
{\cal K}(z,z^\prime) \sim \eta\,\delta(z-z^\prime),
\end{equation}
where  $\eta$ is the bulk viscosity of the fluid.
Under these assumptions only the local equilibrium terms survive and the \hbox{thermo-osmotic} velocity reduces to  $u^x(z)$ given, for 
$z < {h}/{2}$, by 
\begin{equation}
u^x(z) = -\frac{\partial_x T}{\eta}\left. \frac{\partial}{\partial T}\right|_p 
\int_0^{\nicefrac{h}{2}} \ud z^\prime \, {\rm Min}(z,z^\prime)\,\Delta p_\mathrm{T}(z^\prime), 
\label{quasiderj}
\end{equation}
where $\Delta p_\mathrm{T}(z)=p_\mathrm{T}(z)-p$ and the derivative is evaluated at fixed bulk pressure.
This result coincides with the solution of the linearised \hbox{Navier-Stokes} equation for an incompressible fluid in the 
presence of a gradient in the tangential pressure given by 
the LE expression \cite{pp}. 
Moreover, Eq.~(\ref{quasiderj}) reduces to the generalisation of Derjaguin's result~\cite{surface_forces_1987} 
recently provided in Ref.~\cite{ganti17} in the context of nonequilibrium thermodynamics, 
where the enthalpy difference $\Delta h(z)= h(z) -\rho(z)h_m$ takes the place 
of the temperature derivative of $\Delta p_\mathrm{T}(z)$.
All the details about~(\ref{quasiderj}) and the continuum limit can be found in the SM.
Finally, the temperature derivative of the pressure tensor has been recently evaluated by numerical simulations~\cite{han,ganti17} 
for a \hbox{Lennard-Jones} fluid.
Use of the numerical results allows to estimate that the \hbox{thermo-osmotic} velocity for hard walls
is opposite to the thermal gradient and of the order of few micrometer per second. 

In the opposite low density limit, where kinetic theories provide a quantitative interpretation of the phenomenon~\cite{kennard_1938kinetic,sone_2000}, our formalism 
is also able to reproduce the known results. Taking the ideal gas limit, i.e. ignoring the interparticle interactions, 
the gas remains homogeneous and isotropic in the \hbox{$z$-direction} also close to the surface 
implying that ${\cal S}_s(z)=0$. 
The dynamic source term ${\cal S}_d(z)$ can be estimated introducing a 
finite relaxation time $\tau$ and retaining only the kinetic contribution to the 
equilibrium average in~(\ref{dynamic}) as
\begin{equation}
\int_0^\tau \ud t 
\sum_i \left < 
\delta\big(\br-\br_i(t)\big) 
\frac{p_i^x p_i^x(t) p_i^z(t)}{m^2}\left [
\frac{p_i^2}{2m} - m h_m \right ] \right >_0.
\label{eq:dinamo_ideale}
\end{equation}
However, as shown in the SM, this term vanishes in our model because the averaged operator is odd in $p^z$ and 
the ballistic kinetics of an ideal gas conserves both the $x$-component of the momentum and the particles' kinetic energy, 
also when scattering at the confining wall takes place. 
A \hbox{non-zero} value of the average in~(\ref{eq:dinamo_ideale}), and accordingly of the creep velocity, can only be obtained if, 
during the scattering at the surface, at least one of these two conservation laws are violated, as already known in the literature~\cite{maxwell_1879}.
The first case corresponds to elastic scattering against rough surfaces whereas the second case can occur due to
inelastic particle-surface collisions.
Inspired by the seminal work by Maxwell~\cite{maxwell_1879} we assume that, after the collision with the surface, 
the outgoing particle loses memory of the magnitude and the direction of its momentum before the impact.
Within this hypothesis the time-correlation functions vanish after the scattering and~(\ref{eq:dinamo_ideale}) can be evaluated analytically,
leading to the following expression for the \hbox{thermo-osmotic} velocity $v_{\infty}$ far from the surface (the derivation is detailed in the SM)
\begin{equation}
v_{\infty} = \frac{3}{4}\,\frac{\eta}{\rho}\,\frac{\partial_x T}{T}=\frac{3}{4}\,k_{\mathrm{B}}T\, \frac{\eta}{p}\,\frac{\partial_x T}{T},
\end{equation}
which coincides with the kinetic theory result originally obtained by Maxwell~\cite{maxwell_1879,kennard_1938kinetic}
and shows how the slip velocity grows at low pressure, as experimentally demonstrated~\cite{sone_2000}. 

In summary, our generalisation of the Linear Response Theory formalism to inhomogeneous systems, applied 
to a simple microscopic model of fluid close to a planar smooth wall, has provided the general, exact, 
expression allowing to evaluate the \hbox{thermo-osmotic} flow. 
The emerging picture turns out to be more
complex than expected on the basis of the previously adopted theoretical approaches, making
use of kinetic theories as regards \hbox{low-pressure} and rarefied gases and macroscopic linear irreversible thermodynamics for the liquid phase.
The resulting velocity profile of the fluid~(\ref{momenth2}) is valid for all regimes and depends 
on both {\it static} and {\it dynamic} equilibrium properties of the system (see Eq.s~(\ref{equ}) and~(\ref{momenth2})): 
These expressions will be useful in the interpretation of future experiments and numerical simulations in 
the whole phase diagram of a fluid.
A preliminary comparison with the existing macroscopic approach by Derjaguin 
shows that it closely resembles one of the two contributions found in our general expression. 
The other, instead, allows to
reproduce the known expressions of the kinetic theory of gases in the appropriate limits. 

Although our result is expressed in terms of quantities, like the tangential pressure near the wall and the heat 
flux, which are not uniquely defined on microscopic grounds, the {\it combination} of these terms 
(see for instance Eq.~(\ref{static})) is indeed {\it independent} of the 
adopted choice, thereby solving the problem posed in Ref.s~\cite{ganti17,Ganti18}. 

Our method is general: The results presented in this Letter can be easily extended to a closed channel, where the 
relevant quantity is the pressure difference between the two ends of the system and can be applied also to other simple geometries,
like the spherical geometry, where it may provide insights on the microscopic mechanism at the basis of thermophoresis. 

We gratefully thank Roberto Piazza for constant encouragement and  illuminating discussions.

\bibliographystyle{apsrev4-1}

\bibliography{bib_prl}

\begin{thebibliography}{48}%
\makeatletter
\providecommand \@ifxundefined [1]{%
 \@ifx{#1\undefined}
}%
\providecommand \@ifnum [1]{%
 \ifnum #1\expandafter \@firstoftwo
 \else \expandafter \@secondoftwo
 \fi
}%
\providecommand \@ifx [1]{%
 \ifx #1\expandafter \@firstoftwo
 \else \expandafter \@secondoftwo
 \fi
}%
\providecommand \natexlab [1]{#1}%
\providecommand \enquote  [1]{``#1''}%
\providecommand \bibnamefont  [1]{#1}%
\providecommand \bibfnamefont [1]{#1}%
\providecommand \citenamefont [1]{#1}%
\providecommand \href@noop [0]{\@secondoftwo}%
\providecommand \href [0]{\begingroup \@sanitize@url \@href}%
\providecommand \@href[1]{\@@startlink{#1}\@@href}%
\providecommand \@@href[1]{\endgroup#1\@@endlink}%
\providecommand \@sanitize@url [0]{\catcode `\\12\catcode `\$12\catcode
  `\&12\catcode `\#12\catcode `\^12\catcode `\_12\catcode `\%12\relax}%
\providecommand \@@startlink[1]{}%
\providecommand \@@endlink[0]{}%
\providecommand \url  [0]{\begingroup\@sanitize@url \@url }%
\providecommand \@url [1]{\endgroup\@href {#1}{\urlprefix }}%
\providecommand \urlprefix  [0]{URL }%
\providecommand \Eprint [0]{\href }%
\providecommand \doibase [0]{http://dx.doi.org/}%
\providecommand \selectlanguage [0]{\@gobble}%
\providecommand \bibinfo  [0]{\@secondoftwo}%
\providecommand \bibfield  [0]{\@secondoftwo}%
\providecommand \translation [1]{[#1]}%
\providecommand \BibitemOpen [0]{}%
\providecommand \bibitemStop [0]{}%
\providecommand \bibitemNoStop [0]{.\EOS\space}%
\providecommand \EOS [0]{\spacefactor3000\relax}%
\providecommand \BibitemShut  [1]{\csname bibitem#1\endcsname}%
\let\auto@bib@innerbib\@empty
\bibitem [{\citenamefont {Landau}\ and\ \citenamefont
  {Lifshitz}(1987)}]{landau_fluid}%
  \BibitemOpen
  \bibfield  {author} {\bibinfo {author} {\bibfnamefont {L.}~\bibnamefont
  {Landau}}\ and\ \bibinfo {author} {\bibfnamefont {E.}~\bibnamefont
  {Lifshitz}},\ }\href@noop {} {\emph {\bibinfo {title} {Fluid Mechanics}}},\
  \bibinfo {edition} {2nd}\ ed.\ (\bibinfo  {publisher} {Pergamon},\ \bibinfo
  {year} {1987})\BibitemShut {NoStop}%
\bibitem [{\citenamefont {Luttinger}(1964)}]{luttinger_64}%
  \BibitemOpen
  \bibfield  {author} {\bibinfo {author} {\bibfnamefont {J.~M.}\ \bibnamefont
  {Luttinger}},\ }\href {\doibase 10.1103/PhysRev.135.A1505} {\bibfield
  {journal} {\bibinfo  {journal} {Phys. Rev.}\ }\textbf {\bibinfo {volume}
  {135}},\ \bibinfo {pages} {A1505} (\bibinfo {year} {1964})}\BibitemShut
  {NoStop}%
\bibitem [{Note1()}]{Note1}%
  \BibitemOpen
  \bibinfo {note} {See Ref.s~\cite
  {feddersen_1972,crookes_radiometer,reynolds_1879,maxwell_1879}. An historical
  account on the controversies related to the physical understanding of the
  behaviour of the radiometer can be found in Ref.s~\cite
  {maxwell_reynolds_radiometer,piazza_2004}.}\BibitemShut {Stop}%
\bibitem [{Note2()}]{Note2}%
  \BibitemOpen
  \bibinfo {note} {Up to our knowledge, the term \hbox {{\protect \it
  thermo-osmosis}} (or thermal osmosis) has been coined by Lippman~\cite
  {lippmann_1907} in the case of liquids. In gases the same phenomenon is
  referred to as {\protect \it thermal creep} or {\protect \it thermal
  transpiration} (after Reynolds).}\BibitemShut {Stop}%
\bibitem [{\citenamefont {Feddersen}(1873)}]{feddersen_1972}%
  \BibitemOpen
  \bibfield  {author} {\bibinfo {author} {\bibfnamefont {W.}~\bibnamefont
  {Feddersen}},\ }\href {\doibase 10.1080/14786447308640898} {\bibfield
  {journal} {\bibinfo  {journal} {Lond. Edinb. Dubl. Phil. Mag.}\ }\textbf
  {\bibinfo {volume} {46}},\ \bibinfo {pages} {55} (\bibinfo {year}
  {1873})}\BibitemShut {NoStop}%
\bibitem [{\citenamefont {Reynolds}(1879)}]{reynolds_1879}%
  \BibitemOpen
  \bibfield  {author} {\bibinfo {author} {\bibfnamefont {O.}~\bibnamefont
  {Reynolds}},\ }\href {\doibase 10.1098/rstl.1879.0078} {\bibfield  {journal}
  {\bibinfo  {journal} {Philos. Trans. R. Soc. London}\ }\textbf {\bibinfo
  {volume} {170}},\ \bibinfo {pages} {727} (\bibinfo {year}
  {1879})}\BibitemShut {NoStop}%
\bibitem [{\citenamefont {Maxwell}(1879)}]{maxwell_1879}%
  \BibitemOpen
  \bibfield  {author} {\bibinfo {author} {\bibfnamefont {J.~C.}\ \bibnamefont
  {Maxwell}},\ }\href {\doibase 10.1098/rstl.1879.0067} {\bibfield  {journal}
  {\bibinfo  {journal} {Philos. Trans. R. Soc. London}\ }\textbf {\bibinfo
  {volume} {170}},\ \bibinfo {pages} {231} (\bibinfo {year}
  {1879})}\BibitemShut {NoStop}%
\bibitem [{\citenamefont {Lippmann}(1907)}]{lippmann_1907}%
  \BibitemOpen
  \bibfield  {author} {\bibinfo {author} {\bibfnamefont {G.}~\bibnamefont
  {Lippmann}},\ }\href@noop {} {\bibfield  {journal} {\bibinfo  {journal} {C.
  R. Acad. Sci.}\ }\textbf {\bibinfo {volume} {145}},\ \bibinfo {pages} {104}
  (\bibinfo {year} {1907})}\BibitemShut {NoStop}%
\bibitem [{\citenamefont {Aubert}(1912)}]{aubert_1912}%
  \BibitemOpen
  \bibfield  {author} {\bibinfo {author} {\bibfnamefont {M.}~\bibnamefont
  {Aubert}},\ }\href@noop {} {\bibfield  {journal} {\bibinfo  {journal} {Ann.
  Chim. Phys.}\ }\textbf {\bibinfo {volume} {26}},\ \bibinfo {pages} {145}
  (\bibinfo {year} {1912})}\BibitemShut {NoStop}%
\bibitem [{\citenamefont {Derjaguin}\ and\ \citenamefont
  {Sidorenkov}(1941)}]{derjaguin_sidorenkov_1941thermoosmosis}%
  \BibitemOpen
  \bibfield  {author} {\bibinfo {author} {\bibfnamefont {B.~V.}\ \bibnamefont
  {Derjaguin}}\ and\ \bibinfo {author} {\bibfnamefont {G.~P.}\ \bibnamefont
  {Sidorenkov}},\ }\href@noop {} {\bibfield  {journal} {\bibinfo  {journal}
  {Dokl. Akad. Nauk SSSR}\ }\textbf {\bibinfo {volume} {32}},\ \bibinfo {pages}
  {622} (\bibinfo {year} {1941})}\BibitemShut {NoStop}%
\bibitem [{\citenamefont {Zheng}(2002)}]{rev_thermo_gas}%
  \BibitemOpen
  \bibfield  {author} {\bibinfo {author} {\bibfnamefont {F.}~\bibnamefont
  {Zheng}},\ }\href {\doibase 10.1016/S0001-8686(01)00067-7} {\bibfield
  {journal} {\bibinfo  {journal} {Adv. Colloid Interface Sci.}\ }\textbf
  {\bibinfo {volume} {97}},\ \bibinfo {pages} {255 } (\bibinfo {year}
  {2002})}\BibitemShut {NoStop}%
\bibitem [{\citenamefont {Piazza}\ and\ \citenamefont {Parola}(2008)}]{pp}%
  \BibitemOpen
  \bibfield  {author} {\bibinfo {author} {\bibfnamefont {R.}~\bibnamefont
  {Piazza}}\ and\ \bibinfo {author} {\bibfnamefont {A.}~\bibnamefont
  {Parola}},\ }\href {http://stacks.iop.org/0953-8984/20/i=15/a=153102}
  {\bibfield  {journal} {\bibinfo  {journal} {J. Phys. Condens. Matter}\
  }\textbf {\bibinfo {volume} {20}},\ \bibinfo {pages} {153102} (\bibinfo
  {year} {2008})}\BibitemShut {NoStop}%
\bibitem [{\citenamefont {W{\"u}rger}(2010)}]{wurger_2010}%
  \BibitemOpen
  \bibfield  {author} {\bibinfo {author} {\bibfnamefont {A.}~\bibnamefont
  {W{\"u}rger}},\ }\href {\doibase 10.1088/0034-4885/73/12/126601} {\bibfield
  {journal} {\bibinfo  {journal} {Rep. Prog. Phys.}\ ,\ \bibinfo {pages}
  {126601}} (\bibinfo {year} {2010})}\BibitemShut {NoStop}%
\bibitem [{\citenamefont {Golestanian}\ \emph {et~al.}(2007)\citenamefont
  {Golestanian}, \citenamefont {Liverpool},\ and\ \citenamefont
  {Ajdari}}]{swimmers_2007}%
  \BibitemOpen
  \bibfield  {author} {\bibinfo {author} {\bibfnamefont {R.}~\bibnamefont
  {Golestanian}}, \bibinfo {author} {\bibfnamefont {T.~B.}\ \bibnamefont
  {Liverpool}}, \ and\ \bibinfo {author} {\bibfnamefont {A.}~\bibnamefont
  {Ajdari}},\ }\href {http://stacks.iop.org/1367-2630/9/i=5/a=126} {\bibfield
  {journal} {\bibinfo  {journal} {New J. Phys.}\ }\textbf {\bibinfo {volume}
  {9}},\ \bibinfo {pages} {126} (\bibinfo {year} {2007})}\BibitemShut {NoStop}%
\bibitem [{\citenamefont {Chr\'etien}\ \emph {et~al.}(2018)\citenamefont
  {Chr\'etien}, \citenamefont {Bénit}, \citenamefont {Ha}, \citenamefont
  {Keipert}, \citenamefont {El-Khoury}, \citenamefont {Chang}, \citenamefont
  {Jastroch}, \citenamefont {Jacobs}, \citenamefont {Rustin},\ and\
  \citenamefont {Rak}}]{mitocondri}%
  \BibitemOpen
  \bibfield  {author} {\bibinfo {author} {\bibfnamefont {D.}~\bibnamefont
  {Chr\'etien}}, \bibinfo {author} {\bibfnamefont {P.}~\bibnamefont {Bénit}},
  \bibinfo {author} {\bibfnamefont {H.-H.}\ \bibnamefont {Ha}}, \bibinfo
  {author} {\bibfnamefont {S.}~\bibnamefont {Keipert}}, \bibinfo {author}
  {\bibfnamefont {R.}~\bibnamefont {El-Khoury}}, \bibinfo {author}
  {\bibfnamefont {Y.-T.}\ \bibnamefont {Chang}}, \bibinfo {author}
  {\bibfnamefont {M.}~\bibnamefont {Jastroch}}, \bibinfo {author}
  {\bibfnamefont {H.~T.}\ \bibnamefont {Jacobs}}, \bibinfo {author}
  {\bibfnamefont {P.}~\bibnamefont {Rustin}}, \ and\ \bibinfo {author}
  {\bibfnamefont {M.}~\bibnamefont {Rak}},\ }\href {\doibase
  10.1371/journal.pbio.2003992} {\bibfield  {journal} {\bibinfo  {journal}
  {PLOS Biol.}\ }\textbf {\bibinfo {volume} {16}},\ \bibinfo {pages} {1}
  (\bibinfo {year} {2018})}\BibitemShut {NoStop}%
\bibitem [{\citenamefont {Sone}(2000)}]{sone_2000}%
  \BibitemOpen
  \bibfield  {author} {\bibinfo {author} {\bibfnamefont {Y.}~\bibnamefont
  {Sone}},\ }\href {\doibase 10.1146/annurev.fluid.32.1.779} {\bibfield
  {journal} {\bibinfo  {journal} {Annu. Rev. Fluid Mech.}\ }\textbf {\bibinfo
  {volume} {32}},\ \bibinfo {pages} {779} (\bibinfo {year} {2000})}\BibitemShut
  {NoStop}%
\bibitem [{\citenamefont {Eastman}(1926)}]{eastman26}%
  \BibitemOpen
  \bibfield  {author} {\bibinfo {author} {\bibfnamefont {E.~D.}\ \bibnamefont
  {Eastman}},\ }\href {\doibase 10.1021/ja01417a004} {\bibfield  {journal}
  {\bibinfo  {journal} {J. Am. Chem. Soc.}\ }\textbf {\bibinfo {volume} {48}},\
  \bibinfo {pages} {1482} (\bibinfo {year} {1926})}\BibitemShut {NoStop}%
\bibitem [{\citenamefont {{de Groot, S.R.}}(1947)}]{degroot47}%
  \BibitemOpen
  \bibfield  {author} {\bibinfo {author} {\bibnamefont {{de Groot, S.R.}}},\
  }\href {\doibase 10.1051/jphysrad:0194700806018801} {\bibfield  {journal}
  {\bibinfo  {journal} {J. Phys. Radium}\ }\textbf {\bibinfo {volume} {8}},\
  \bibinfo {pages} {188} (\bibinfo {year} {1947})}\BibitemShut {NoStop}%
\bibitem [{\citenamefont {Denbigh}\ and\ \citenamefont
  {Raumann}(1952)}]{denbigh1952}%
  \BibitemOpen
  \bibfield  {author} {\bibinfo {author} {\bibfnamefont {K.~G.}\ \bibnamefont
  {Denbigh}}\ and\ \bibinfo {author} {\bibfnamefont {G.}~\bibnamefont
  {Raumann}},\ }\href {\doibase 10.1098/rspa.1952.0007} {\bibfield  {journal}
  {\bibinfo  {journal} {Proc. R. Soc. London, Ser. A}\ }\textbf {\bibinfo
  {volume} {210}},\ \bibinfo {pages} {377} (\bibinfo {year}
  {1952})}\BibitemShut {NoStop}%
\bibitem [{\citenamefont {Hutchison}\ \emph {et~al.}(1948)\citenamefont
  {Hutchison}, \citenamefont {Nixon},\ and\ \citenamefont
  {Denbigh}}]{hutchisonnixondenbigh_1948}%
  \BibitemOpen
  \bibfield  {author} {\bibinfo {author} {\bibfnamefont {H.~P.}\ \bibnamefont
  {Hutchison}}, \bibinfo {author} {\bibfnamefont {I.~S.}\ \bibnamefont
  {Nixon}}, \ and\ \bibinfo {author} {\bibfnamefont {K.~G.}\ \bibnamefont
  {Denbigh}},\ }\href {\doibase 10.1039/DF9480300086} {\bibfield  {journal}
  {\bibinfo  {journal} {Discuss. Faraday Soc.}\ }\textbf {\bibinfo {volume}
  {3}},\ \bibinfo {pages} {86} (\bibinfo {year} {1948})}\BibitemShut {NoStop}%
\bibitem [{\citenamefont {Pagliuca}\ \emph {et~al.}(1987)\citenamefont
  {Pagliuca}, \citenamefont {Bencivenga}, \citenamefont {Mita}, \citenamefont
  {Perna},\ and\ \citenamefont {Gaeta}}]{gaeta}%
  \BibitemOpen
  \bibfield  {author} {\bibinfo {author} {\bibfnamefont {N.}~\bibnamefont
  {Pagliuca}}, \bibinfo {author} {\bibfnamefont {U.}~\bibnamefont
  {Bencivenga}}, \bibinfo {author} {\bibfnamefont {D.}~\bibnamefont {Mita}},
  \bibinfo {author} {\bibfnamefont {G.}~\bibnamefont {Perna}}, \ and\ \bibinfo
  {author} {\bibfnamefont {F.}~\bibnamefont {Gaeta}},\ }\href {\doibase
  https://doi.org/10.1016/S0376-7388(00)80049-6} {\bibfield  {journal}
  {\bibinfo  {journal} {J. Mem. Sci.}\ }\textbf {\bibinfo {volume} {33}},\
  \bibinfo {pages} {1} (\bibinfo {year} {1987})}\BibitemShut {NoStop}%
\bibitem [{\citenamefont {Bregulla}\ \emph {et~al.}(2016)\citenamefont
  {Bregulla}, \citenamefont {W\"urger}, \citenamefont {G\"unther},
  \citenamefont {Mertig},\ and\ \citenamefont {Cichos}}]{bregulla_2016}%
  \BibitemOpen
  \bibfield  {author} {\bibinfo {author} {\bibfnamefont {A.~P.}\ \bibnamefont
  {Bregulla}}, \bibinfo {author} {\bibfnamefont {A.}~\bibnamefont {W\"urger}},
  \bibinfo {author} {\bibfnamefont {K.}~\bibnamefont {G\"unther}}, \bibinfo
  {author} {\bibfnamefont {M.}~\bibnamefont {Mertig}}, \ and\ \bibinfo {author}
  {\bibfnamefont {F.}~\bibnamefont {Cichos}},\ }\href {\doibase
  10.1103/PhysRevLett.116.188303} {\bibfield  {journal} {\bibinfo  {journal}
  {Phys. Rev. Lett.}\ }\textbf {\bibinfo {volume} {116}},\ \bibinfo {pages}
  {188303} (\bibinfo {year} {2016})}\BibitemShut {NoStop}%
\bibitem [{\citenamefont {Barrag{\'a}n}\ and\ \citenamefont
  {Kjelstrup}(2017)}]{barragankjelstrup_2016review}%
  \BibitemOpen
  \bibfield  {author} {\bibinfo {author} {\bibfnamefont {V.~M.}\ \bibnamefont
  {Barrag{\'a}n}}\ and\ \bibinfo {author} {\bibfnamefont {S.}~\bibnamefont
  {Kjelstrup}},\ }\href {\doibase 10.1515/jnet-2016-0088} {\bibfield  {journal}
  {\bibinfo  {journal} {J. Non-Equilib. Thermodyn.}\ }\textbf {\bibinfo
  {volume} {42}},\ \bibinfo {pages} {217} (\bibinfo {year} {2017})}\BibitemShut
  {NoStop}%
\bibitem [{\citenamefont {Derjaguin}\ \emph {et~al.}(1987)\citenamefont
  {Derjaguin}, \citenamefont {Churaev},\ and\ \citenamefont
  {Muller}}]{surface_forces_1987}%
  \BibitemOpen
  \bibfield  {author} {\bibinfo {author} {\bibfnamefont {B.~V.}\ \bibnamefont
  {Derjaguin}}, \bibinfo {author} {\bibfnamefont {N.~V.}\ \bibnamefont
  {Churaev}}, \ and\ \bibinfo {author} {\bibfnamefont {V.~M.}\ \bibnamefont
  {Muller}},\ }\enquote {\bibinfo {title} {{Surface Forces in Transport
  Phenomena}},}\ in\ \href {\doibase 10.1007/978-1-4757-6639-4_11} {\emph
  {\bibinfo {booktitle} {{Surface Forces}}}}\ (\bibinfo  {publisher}
  {Springer},\ \bibinfo {address} {Boston},\ \bibinfo {year}
  {1987})\BibitemShut {NoStop}%
\bibitem [{\citenamefont {Anderson}(1989)}]{anderson_review}%
  \BibitemOpen
  \bibfield  {author} {\bibinfo {author} {\bibfnamefont {J.~L.}\ \bibnamefont
  {Anderson}},\ }\href {\doibase 10.1146/annurev.fl.21.010189.000425}
  {\bibfield  {journal} {\bibinfo  {journal} {Annu. Rev. Fluid Mech.}\ }\textbf
  {\bibinfo {volume} {21}},\ \bibinfo {pages} {61} (\bibinfo {year}
  {1989})}\BibitemShut {NoStop}%
\bibitem [{\citenamefont {Han}(2005)}]{han}%
  \BibitemOpen
  \bibfield  {author} {\bibinfo {author} {\bibfnamefont {M.}~\bibnamefont
  {Han}},\ }\href {\doibase 10.1016/j.jcis.2004.09.067} {\bibfield  {journal}
  {\bibinfo  {journal} {J. Colloid Interface Sci.}\ }\textbf {\bibinfo {volume}
  {284}},\ \bibinfo {pages} {339—348} (\bibinfo {year} {2005})}\BibitemShut
  {NoStop}%
\bibitem [{\citenamefont {Wold}\ and\ \citenamefont
  {Hafskjold}(1999)}]{bjorn1999}%
  \BibitemOpen
  \bibfield  {author} {\bibinfo {author} {\bibfnamefont {I.}~\bibnamefont
  {Wold}}\ and\ \bibinfo {author} {\bibfnamefont {B.}~\bibnamefont
  {Hafskjold}},\ }\href {\doibase 10.1023/A:1022631102246} {\bibfield
  {journal} {\bibinfo  {journal} {Int. J. Thermophys.}\ }\textbf {\bibinfo
  {volume} {20}},\ \bibinfo {pages} {847} (\bibinfo {year} {1999})}\BibitemShut
  {NoStop}%
\bibitem [{\citenamefont {Galli{\'e}ro}\ \emph {et~al.}(2002)\citenamefont
  {Galli{\'e}ro}, \citenamefont {Colombani}, \citenamefont {Duguay},
  \citenamefont {Caltagirone},\ and\ \citenamefont {Montel}}]{galliero2002}%
  \BibitemOpen
  \bibfield  {author} {\bibinfo {author} {\bibfnamefont {G.}~\bibnamefont
  {Galli{\'e}ro}}, \bibinfo {author} {\bibfnamefont {J.}~\bibnamefont
  {Colombani}}, \bibinfo {author} {\bibfnamefont {B.}~\bibnamefont {Duguay}},
  \bibinfo {author} {\bibfnamefont {J.-P.}\ \bibnamefont {Caltagirone}}, \ and\
  \bibinfo {author} {\bibfnamefont {F.}~\bibnamefont {Montel}},\ }\href@noop {}
  {\bibfield  {journal} {\bibinfo  {journal} {Entropie}\ }\textbf {\bibinfo
  {volume} {239/240}},\ \bibinfo {pages} {98} (\bibinfo {year}
  {2002})}\BibitemShut {NoStop}%
\bibitem [{\citenamefont {Ganti}\ \emph {et~al.}(2017)\citenamefont {Ganti},
  \citenamefont {Liu},\ and\ \citenamefont {Frenkel}}]{ganti17}%
  \BibitemOpen
  \bibfield  {author} {\bibinfo {author} {\bibfnamefont {R.}~\bibnamefont
  {Ganti}}, \bibinfo {author} {\bibfnamefont {Y.}~\bibnamefont {Liu}}, \ and\
  \bibinfo {author} {\bibfnamefont {D.}~\bibnamefont {Frenkel}},\ }\href
  {\doibase 10.1103/PhysRevLett.119.038002} {\bibfield  {journal} {\bibinfo
  {journal} {Phys. Rev. Lett.}\ }\textbf {\bibinfo {volume} {119}},\ \bibinfo
  {pages} {038002} (\bibinfo {year} {2017})}\BibitemShut {NoStop}%
\bibitem [{\citenamefont {Ganti}\ \emph {et~al.}(2018)\citenamefont {Ganti},
  \citenamefont {Liu},\ and\ \citenamefont {Frenkel}}]{Ganti18}%
  \BibitemOpen
  \bibfield  {author} {\bibinfo {author} {\bibfnamefont {R.}~\bibnamefont
  {Ganti}}, \bibinfo {author} {\bibfnamefont {Y.}~\bibnamefont {Liu}}, \ and\
  \bibinfo {author} {\bibfnamefont {D.}~\bibnamefont {Frenkel}},\ }\href
  {\doibase 10.1103/PhysRevLett.121.068002} {\bibfield  {journal} {\bibinfo
  {journal} {Phys. Rev. Lett.}\ }\textbf {\bibinfo {volume} {121}},\ \bibinfo
  {pages} {068002} (\bibinfo {year} {2018})}\BibitemShut {NoStop}%
\bibitem [{\citenamefont {Fu}\ \emph {et~al.}(2017)\citenamefont {Fu},
  \citenamefont {Merabia},\ and\ \citenamefont {Joly}}]{joly17}%
  \BibitemOpen
  \bibfield  {author} {\bibinfo {author} {\bibfnamefont {L.}~\bibnamefont
  {Fu}}, \bibinfo {author} {\bibfnamefont {S.}~\bibnamefont {Merabia}}, \ and\
  \bibinfo {author} {\bibfnamefont {L.}~\bibnamefont {Joly}},\ }\href {\doibase
  10.1103/PhysRevLett.119.214501} {\bibfield  {journal} {\bibinfo  {journal}
  {Phys. Rev. Lett.}\ }\textbf {\bibinfo {volume} {119}},\ \bibinfo {pages}
  {214501} (\bibinfo {year} {2017})}\BibitemShut {NoStop}%
\bibitem [{\citenamefont {Fu}\ \emph {et~al.}(2018)\citenamefont {Fu},
  \citenamefont {Merabia},\ and\ \citenamefont {Joly}}]{joly18}%
  \BibitemOpen
  \bibfield  {author} {\bibinfo {author} {\bibfnamefont {L.}~\bibnamefont
  {Fu}}, \bibinfo {author} {\bibfnamefont {S.}~\bibnamefont {Merabia}}, \ and\
  \bibinfo {author} {\bibfnamefont {L.}~\bibnamefont {Joly}},\ }\href {\doibase
  10.1021/acs.jpclett.8b00703} {\bibfield  {journal} {\bibinfo  {journal} {J.
  Phys. Chem. Lett.}\ }\textbf {\bibinfo {volume} {9}},\ \bibinfo {pages}
  {2086} (\bibinfo {year} {2018})},\ \bibinfo {note} {pMID:
  29624390}\BibitemShut {NoStop}%
\bibitem [{\citenamefont {Kubo}(1957)}]{kubo}%
  \BibitemOpen
  \bibfield  {author} {\bibinfo {author} {\bibfnamefont {R.}~\bibnamefont
  {Kubo}},\ }\href {\doibase 10.1143/JPSJ.12.570} {\bibfield  {journal}
  {\bibinfo  {journal} {J. Phys. Soc. Jpn.}\ }\textbf {\bibinfo {volume}
  {12}},\ \bibinfo {pages} {570} (\bibinfo {year} {1957})}\BibitemShut
  {NoStop}%
\bibitem [{\citenamefont {Zwanzig}(1965)}]{zwanzig}%
  \BibitemOpen
  \bibfield  {author} {\bibinfo {author} {\bibfnamefont {R.}~\bibnamefont
  {Zwanzig}},\ }\href {\doibase 10.1146/annurev.pc.16.100165.000435} {\bibfield
   {journal} {\bibinfo  {journal} {Annu. Rev. Phys. Chem.}\ }\textbf {\bibinfo
  {volume} {16}},\ \bibinfo {pages} {67} (\bibinfo {year} {1965})}\BibitemShut
  {NoStop}%
\bibitem [{\citenamefont {Mori}(1956)}]{mori56}%
  \BibitemOpen
  \bibfield  {author} {\bibinfo {author} {\bibfnamefont {H.}~\bibnamefont
  {Mori}},\ }\href {\doibase 10.1143/JPSJ.11.1029} {\bibfield  {journal}
  {\bibinfo  {journal} {J. Phys. Soc. Jpn.}\ }\textbf {\bibinfo {volume}
  {11}},\ \bibinfo {pages} {1029} (\bibinfo {year} {1956})}\BibitemShut
  {NoStop}%
\bibitem [{\citenamefont {Mori}(1958)}]{mori58}%
  \BibitemOpen
  \bibfield  {author} {\bibinfo {author} {\bibfnamefont {H.}~\bibnamefont
  {Mori}},\ }\href {\doibase 10.1103/PhysRev.112.1829} {\bibfield  {journal}
  {\bibinfo  {journal} {Phys. Rev.}\ }\textbf {\bibinfo {volume} {112}},\
  \bibinfo {pages} {1829} (\bibinfo {year} {1958})}\BibitemShut {NoStop}%
\bibitem [{Note99()}]{Note99}%
  \BibitemOpen
  \bibinfo {note} {\label {foot:definizione_flussi}Note that the conservation
  equations just define the divergence of the current operators. In some case,
  notably for the pressure tensor and for the heat flux, this fact poses some
  mathematical ambiguity in the precise definition of the current as discussed
  in detail elsewhere (see Ref.s~\cite {henderson,baroni_15} and Supplementary
  Material). In the following we will just assume that the adopted definition
  of the current preserves the short range nature of the dynamic correlations
  and the spatial symmetries of the system.}\BibitemShut {Stop}%
\bibitem [{\citenamefont {Balescu}(1975)}]{balescu}%
  \BibitemOpen
  \bibfield  {author} {\bibinfo {author} {\bibfnamefont {R.}~\bibnamefont
  {Balescu}},\ }\href@noop {} {\emph {\bibinfo {title} {{Equilibrium and
  nonequilibrium statistical mechanics}}}}\ (\bibinfo  {publisher} {John Wiley
  \& Sons},\ \bibinfo {year} {1975})\BibitemShut {NoStop}%
\bibitem [{Note100()}]{Note100}%
  \BibitemOpen
  \bibinfo {note} {Special care must be paid in taking the \hbox {long-time}
  limit for correlation functions involving currents of conserved quantities,
  as in our cases~\cite {simpleliquids_fourth}.}\BibitemShut {Stop}%
\bibitem [{\citenamefont {{Hansen, Jean-Pierre and McDonald, Ian
  R.}}(2013)}]{simpleliquids_fourth}%
  \BibitemOpen
  \bibfield  {author} {\bibinfo {author} {\bibnamefont {{Hansen, Jean-Pierre
  and McDonald, Ian R.}}},\ }\href@noop {} {\emph {\bibinfo {title} {{Theory of
  Simple Liquids}}}},\ \bibinfo {edition} {4th}\ ed.\ (\bibinfo  {publisher}
  {Academic Press},\ \bibinfo {year} {2013})\BibitemShut {NoStop}%
\bibitem [{Note101()}]{Note101}%
  \BibitemOpen
  \bibinfo {note} {The first contribution to the integral in Eq.~(\ref
  {momenth2}) vanishes in a homogeneous fluid due to the independence of heat
  and mass current fluctuations. See e.g. Ref.~\cite {landau_fluid}, Sec. 49 or
  Ref.~\cite {balescu}, Sec. 12.5.}\BibitemShut {Stop}%
\bibitem [{\citenamefont {Vanossi}\ \emph {et~al.}(2013)\citenamefont
  {Vanossi}, \citenamefont {Manini}, \citenamefont {Urbakh}, \citenamefont
  {Zapperi},\ and\ \citenamefont {Tosatti}}]{friction}%
  \BibitemOpen
  \bibfield  {author} {\bibinfo {author} {\bibfnamefont {A.}~\bibnamefont
  {Vanossi}}, \bibinfo {author} {\bibfnamefont {N.}~\bibnamefont {Manini}},
  \bibinfo {author} {\bibfnamefont {M.}~\bibnamefont {Urbakh}}, \bibinfo
  {author} {\bibfnamefont {S.}~\bibnamefont {Zapperi}}, \ and\ \bibinfo
  {author} {\bibfnamefont {E.}~\bibnamefont {Tosatti}},\ }\href {\doibase
  10.1103/RevModPhys.85.529} {\bibfield  {journal} {\bibinfo  {journal} {Rev.
  Mod. Phys.}\ }\textbf {\bibinfo {volume} {85}},\ \bibinfo {pages} {529}
  (\bibinfo {year} {2013})}\BibitemShut {NoStop}%
\bibitem [{\citenamefont {Kennard}(1938)}]{kennard_1938kinetic}%
  \BibitemOpen
  \bibfield  {author} {\bibinfo {author} {\bibfnamefont {E.}~\bibnamefont
  {Kennard}},\ }\href@noop {} {\emph {\bibinfo {title} {Kinetic theory of
  gases: with an introduction to statistical mechanics}}}\ (\bibinfo
  {publisher} {McGraw-Hill},\ \bibinfo {year} {1938})\BibitemShut {NoStop}%
\bibitem [{\citenamefont {Woodruff}(1966)}]{crookes_radiometer}%
  \BibitemOpen
  \bibfield  {author} {\bibinfo {author} {\bibfnamefont {A.~E.}\ \bibnamefont
  {Woodruff}},\ }\href {http://www.jstor.org/stable/227958} {\bibfield
  {journal} {\bibinfo  {journal} {Isis}\ }\textbf {\bibinfo {volume} {57}},\
  \bibinfo {pages} {188} (\bibinfo {year} {1966})}\BibitemShut {NoStop}%
\bibitem [{\citenamefont {Brush}\ and\ \citenamefont
  {Everitt}(1969)}]{maxwell_reynolds_radiometer}%
  \BibitemOpen
  \bibfield  {author} {\bibinfo {author} {\bibfnamefont {S.~G.}\ \bibnamefont
  {Brush}}\ and\ \bibinfo {author} {\bibfnamefont {C.~W.~F.}\ \bibnamefont
  {Everitt}},\ }\href {http://www.jstor.org/stable/27757296} {\bibfield
  {journal} {\bibinfo  {journal} {Hist. St. Phys. Sci.}\ }\textbf {\bibinfo
  {volume} {1}},\ \bibinfo {pages} {105} (\bibinfo {year} {1969})}\BibitemShut
  {NoStop}%
\bibitem [{\citenamefont {Piazza}(2004)}]{piazza_2004}%
  \BibitemOpen
  \bibfield  {author} {\bibinfo {author} {\bibfnamefont {R.}~\bibnamefont
  {Piazza}},\ }\href {http://stacks.iop.org/0953-8984/16/i=38/a=032} {\bibfield
   {journal} {\bibinfo  {journal} {J. Phys. Condens. Matter}\ }\textbf
  {\bibinfo {volume} {16}},\ \bibinfo {pages} {S4195} (\bibinfo {year}
  {2004})}\BibitemShut {NoStop}%
\bibitem [{\citenamefont {Schofield}\ and\ \citenamefont
  {Henderson}(1982)}]{henderson}%
  \BibitemOpen
  \bibfield  {author} {\bibinfo {author} {\bibfnamefont {P.}~\bibnamefont
  {Schofield}}\ and\ \bibinfo {author} {\bibfnamefont {J.~R.}\ \bibnamefont
  {Henderson}},\ }\href {\doibase 10.1098/rspa.1982.0015} {\bibfield  {journal}
  {\bibinfo  {journal} {Proc. R. Soc. London, Ser. A}\ }\textbf {\bibinfo
  {volume} {379}},\ \bibinfo {pages} {231} (\bibinfo {year}
  {1982})}\BibitemShut {NoStop}%
\bibitem [{\citenamefont {Ercole}\ \emph {et~al.}(2016)\citenamefont {Ercole},
  \citenamefont {Marcolongo}, \citenamefont {Umari},\ and\ \citenamefont
  {Baroni}}]{baroni_15}%
  \BibitemOpen
  \bibfield  {author} {\bibinfo {author} {\bibfnamefont {L.}~\bibnamefont
  {Ercole}}, \bibinfo {author} {\bibfnamefont {A.}~\bibnamefont {Marcolongo}},
  \bibinfo {author} {\bibfnamefont {P.}~\bibnamefont {Umari}}, \ and\ \bibinfo
  {author} {\bibfnamefont {S.}~\bibnamefont {Baroni}},\ }\href {\doibase
  10.1007/s10909-016-1617-6} {\bibfield  {journal} {\bibinfo  {journal} {J. Low
  Temp. Phys.}\ }\textbf {\bibinfo {volume} {185}},\ \bibinfo {pages} {79}
  (\bibinfo {year} {2016})}\BibitemShut {NoStop}%
\end{thebibliography}%


\begin{thebibliography}{21}%
\makeatletter
\providecommand \@ifxundefined [1]{%
 \@ifx{#1\undefined}
}%
\providecommand \@ifnum [1]{%
 \ifnum #1\expandafter \@firstoftwo
 \else \expandafter \@secondoftwo
 \fi
}%
\providecommand \@ifx [1]{%
 \ifx #1\expandafter \@firstoftwo
 \else \expandafter \@secondoftwo
 \fi
}%
\providecommand \natexlab [1]{#1}%
\providecommand \enquote  [1]{``#1''}%
\providecommand \bibnamefont  [1]{#1}%
\providecommand \bibfnamefont [1]{#1}%
\providecommand \citenamefont [1]{#1}%
\providecommand \href@noop [0]{\@secondoftwo}%
\providecommand \href [0]{\begingroup \@sanitize@url \@href}%
\providecommand \@href[1]{\@@startlink{#1}\@@href}%
\providecommand \@@href[1]{\endgroup#1\@@endlink}%
\providecommand \@sanitize@url [0]{\catcode `\\12\catcode `\$12\catcode
  `\&12\catcode `\#12\catcode `\^12\catcode `\_12\catcode `\%12\relax}%
\providecommand \@@startlink[1]{}%
\providecommand \@@endlink[0]{}%
\providecommand \url  [0]{\begingroup\@sanitize@url \@url }%
\providecommand \@url [1]{\endgroup\@href {#1}{\urlprefix }}%
\providecommand \urlprefix  [0]{URL }%
\providecommand \Eprint [0]{\href }%
\providecommand \doibase [0]{http://dx.doi.org/}%
\providecommand \selectlanguage [0]{\@gobble}%
\providecommand \bibinfo  [0]{\@secondoftwo}%
\providecommand \bibfield  [0]{\@secondoftwo}%
\providecommand \translation [1]{[#1]}%
\providecommand \BibitemOpen [0]{}%
\providecommand \bibitemStop [0]{}%
\providecommand \bibitemNoStop [0]{.\EOS\space}%
\providecommand \EOS [0]{\spacefactor3000\relax}%
\providecommand \BibitemShut  [1]{\csname bibitem#1\endcsname}%
\let\auto@bib@innerbib\@empty
\bibitem [{\citenamefont {Rowlinson}(1993)}]{rowlinson_1993}%
  \BibitemOpen
  \bibfield  {author} {\bibinfo {author} {\bibfnamefont {J.~S.}\ \bibnamefont
  {Rowlinson}},\ }\href {\doibase https://doi.org/10.1006/jcht.1993.1154}
  {\bibfield  {journal} {\bibinfo  {journal} {The Journal of Chemical
  Thermodynamics}\ }\textbf {\bibinfo {volume} {25}},\ \bibinfo {pages} {449 }
  (\bibinfo {year} {1993})}\BibitemShut {NoStop}%
\bibitem [{\citenamefont {Irving}\ and\ \citenamefont
  {Kirkwood}(1950)}]{irvingkirkwood_1949}%
  \BibitemOpen
  \bibfield  {author} {\bibinfo {author} {\bibfnamefont {J.~H.}\ \bibnamefont
  {Irving}}\ and\ \bibinfo {author} {\bibfnamefont {J.~G.}\ \bibnamefont
  {Kirkwood}},\ }\href {\doibase 10.1063/1.1747782} {\bibfield  {journal}
  {\bibinfo  {journal} {The Journal of Chemical Physics}\ }\textbf {\bibinfo
  {volume} {18}},\ \bibinfo {pages} {817} (\bibinfo {year} {1950})}\BibitemShut
  {NoStop}%
\bibitem [{\citenamefont {Ercole}\ \emph {et~al.}(2016)\citenamefont {Ercole},
  \citenamefont {Marcolongo}, \citenamefont {Umari},\ and\ \citenamefont
  {Baroni}}]{baroni_15}%
  \BibitemOpen
  \bibfield  {author} {\bibinfo {author} {\bibfnamefont {L.}~\bibnamefont
  {Ercole}}, \bibinfo {author} {\bibfnamefont {A.}~\bibnamefont {Marcolongo}},
  \bibinfo {author} {\bibfnamefont {P.}~\bibnamefont {Umari}}, \ and\ \bibinfo
  {author} {\bibfnamefont {S.}~\bibnamefont {Baroni}},\ }\href {\doibase
  10.1007/s10909-016-1617-6} {\bibfield  {journal} {\bibinfo  {journal}
  {Journal of Low Temperature Physics}\ }\textbf {\bibinfo {volume} {185}},\
  \bibinfo {pages} {79} (\bibinfo {year} {2016})}\BibitemShut {NoStop}%
\bibitem [{\citenamefont {Balescu}(1975)}]{balescu_equilibrium_1975}%
  \BibitemOpen
  \bibfield  {author} {\bibinfo {author} {\bibfnamefont {R.}~\bibnamefont
  {Balescu}},\ }\href@noop {} {\emph {\bibinfo {title} {{Equilibrium and
  nonequilibrium statistical mechanics}}}}\ (\bibinfo  {publisher} {John Wiley
  \& Sons},\ \bibinfo {year} {1975})\BibitemShut {NoStop}%
\bibitem [{\citenamefont {Schofield}\ and\ \citenamefont
  {Henderson}(1982)}]{schofield_henderson_1982}%
  \BibitemOpen
  \bibfield  {author} {\bibinfo {author} {\bibfnamefont {P.}~\bibnamefont
  {Schofield}}\ and\ \bibinfo {author} {\bibfnamefont {J.~R.}\ \bibnamefont
  {Henderson}},\ }\href {\doibase 10.1098/rspa.1982.0015} {\bibfield  {journal}
  {\bibinfo  {journal} {Proceedings of the Royal Society of London A}\ }\textbf
  {\bibinfo {volume} {379}},\ \bibinfo {pages} {231} (\bibinfo {year}
  {1982})}\BibitemShut {NoStop}%
\bibitem [{\citenamefont {{Hansen, Jean-Pierre and McDonald, Ian
  R.}}(2013)}]{simpleliquids_fourth}%
  \BibitemOpen
  \bibfield  {author} {\bibinfo {author} {\bibnamefont {{Hansen, Jean-Pierre
  and McDonald, Ian R.}}},\ }\href@noop {} {\emph {\bibinfo {title} {{Theory of
  Simple Liquids}}}},\ \bibinfo {edition} {4th}\ ed.\ (\bibinfo  {publisher}
  {Academic Press},\ \bibinfo {year} {2013})\BibitemShut {NoStop}%
\bibitem [{\citenamefont {Kirkwood}\ and\ \citenamefont
  {Buff}(1949)}]{kirkwoodbuff_1949}%
  \BibitemOpen
  \bibfield  {author} {\bibinfo {author} {\bibfnamefont {J.~G.}\ \bibnamefont
  {Kirkwood}}\ and\ \bibinfo {author} {\bibfnamefont {F.~P.}\ \bibnamefont
  {Buff}},\ }\href {\doibase 10.1063/1.1747248} {\bibfield  {journal} {\bibinfo
   {journal} {The Journal of Chemical Physics}\ }\textbf {\bibinfo {volume}
  {17}},\ \bibinfo {pages} {338} (\bibinfo {year} {1949})}\BibitemShut
  {NoStop}%
\bibitem [{\citenamefont {Harasima}(1958)}]{harasima_1958}%
  \BibitemOpen
  \bibfield  {author} {\bibinfo {author} {\bibfnamefont {A.}~\bibnamefont
  {Harasima}},\ }\href@noop {} {\bibfield  {journal} {\bibinfo  {journal}
  {Advances in Chemical Physics}\ }\textbf {\bibinfo {volume} {1}},\ \bibinfo
  {pages} {203 } (\bibinfo {year} {1958})}\BibitemShut {NoStop}%
\bibitem [{\citenamefont {Ono}\ and\ \citenamefont
  {Kondo}(1960)}]{ono_kondo_60}%
  \BibitemOpen
  \bibfield  {author} {\bibinfo {author} {\bibfnamefont {S.}~\bibnamefont
  {Ono}}\ and\ \bibinfo {author} {\bibfnamefont {S.}~\bibnamefont {Kondo}},\
  }\enquote {\bibinfo {title} {Molecular theory of surface tension in
  liquids},}\ in\ \href {\doibase 10.1007/978-3-642-45947-4_2} {\emph {\bibinfo
  {booktitle} {Structure of Liquids}}}\ (\bibinfo  {publisher} {Springer Berlin
  Heidelberg},\ \bibinfo {address} {Berlin, Heidelberg},\ \bibinfo {year}
  {1960})\ pp.\ \bibinfo {pages} {134--280}\BibitemShut {NoStop}%
\bibitem [{\citenamefont {Baus}\ and\ \citenamefont
  {Lovett}(1990)}]{baus_lovettPRL90}%
  \BibitemOpen
  \bibfield  {author} {\bibinfo {author} {\bibfnamefont {M.}~\bibnamefont
  {Baus}}\ and\ \bibinfo {author} {\bibfnamefont {R.}~\bibnamefont {Lovett}},\
  }\href {\doibase 10.1103/PhysRevLett.65.1781} {\bibfield  {journal} {\bibinfo
   {journal} {Physical Review Letters}\ }\textbf {\bibinfo {volume} {65}},\
  \bibinfo {pages} {1781} (\bibinfo {year} {1990})}\BibitemShut {NoStop}%
\bibitem [{\citenamefont {Baus}\ and\ \citenamefont
  {Lovett}(1991)}]{baus_lovettPRA91}%
  \BibitemOpen
  \bibfield  {author} {\bibinfo {author} {\bibfnamefont {M.}~\bibnamefont
  {Baus}}\ and\ \bibinfo {author} {\bibfnamefont {R.}~\bibnamefont {Lovett}},\
  }\href {\doibase 10.1103/PhysRevA.44.1211} {\bibfield  {journal} {\bibinfo
  {journal} {Physical Review A}\ }\textbf {\bibinfo {volume} {44}},\ \bibinfo
  {pages} {1211} (\bibinfo {year} {1991})}\BibitemShut {NoStop}%
\bibitem [{\citenamefont {Rowlinson}(1991)}]{rowlinson_replyPRL}%
  \BibitemOpen
  \bibfield  {author} {\bibinfo {author} {\bibfnamefont {J.~S.}\ \bibnamefont
  {Rowlinson}},\ }\href {\doibase 10.1103/PhysRevLett.67.406} {\bibfield
  {journal} {\bibinfo  {journal} {Physical Review Letters}\ }\textbf {\bibinfo
  {volume} {67}},\ \bibinfo {pages} {406} (\bibinfo {year} {1991})}\BibitemShut
  {NoStop}%
\bibitem [{\citenamefont {Ganti}\ \emph {et~al.}(2017)\citenamefont {Ganti},
  \citenamefont {Liu},\ and\ \citenamefont {Frenkel}}]{ganti_2017}%
  \BibitemOpen
  \bibfield  {author} {\bibinfo {author} {\bibfnamefont {R.}~\bibnamefont
  {Ganti}}, \bibinfo {author} {\bibfnamefont {Y.}~\bibnamefont {Liu}}, \ and\
  \bibinfo {author} {\bibfnamefont {D.}~\bibnamefont {Frenkel}},\ }\href
  {\doibase 10.1103/PhysRevLett.119.038002} {\bibfield  {journal} {\bibinfo
  {journal} {Physical Review Letters}\ }\textbf {\bibinfo {volume} {119}},\
  \bibinfo {pages} {038002} (\bibinfo {year} {2017})}\BibitemShut {NoStop}%
\bibitem [{\citenamefont {Ganti}\ \emph {et~al.}(2018)\citenamefont {Ganti},
  \citenamefont {Liu},\ and\ \citenamefont {Frenkel}}]{ganti_2018}%
  \BibitemOpen
  \bibfield  {author} {\bibinfo {author} {\bibfnamefont {R.}~\bibnamefont
  {Ganti}}, \bibinfo {author} {\bibfnamefont {Y.}~\bibnamefont {Liu}}, \ and\
  \bibinfo {author} {\bibfnamefont {D.}~\bibnamefont {Frenkel}},\ }\href
  {\doibase 10.1103/PhysRevLett.121.068002} {\bibfield  {journal} {\bibinfo
  {journal} {{Physical Review Letters}}\ }\textbf {\bibinfo {volume} {121}},\
  \bibinfo {pages} {068002} (\bibinfo {year} {2018})}\BibitemShut {NoStop}%
\bibitem [{\citenamefont {Mori}(1956)}]{mori_1956}%
  \BibitemOpen
  \bibfield  {author} {\bibinfo {author} {\bibfnamefont {H.}~\bibnamefont
  {Mori}},\ }\href {\doibase 10.1143/JPSJ.11.1029} {\bibfield  {journal}
  {\bibinfo  {journal} {Journal of the Physical Society of Japan}\ }\textbf
  {\bibinfo {volume} {11}},\ \bibinfo {pages} {1029} (\bibinfo {year}
  {1956})}\BibitemShut {NoStop}%
\bibitem [{\citenamefont {Mori}(1958)}]{mori_1958}%
  \BibitemOpen
  \bibfield  {author} {\bibinfo {author} {\bibfnamefont {H.}~\bibnamefont
  {Mori}},\ }\href {\doibase 10.1103/PhysRev.112.1829} {\bibfield  {journal}
  {\bibinfo  {journal} {Physical Review}\ }\textbf {\bibinfo {volume} {112}},\
  \bibinfo {pages} {1829} (\bibinfo {year} {1958})}\BibitemShut {NoStop}%
\bibitem [{\citenamefont {Derjaguin}\ \emph {et~al.}(1987)\citenamefont
  {Derjaguin}, \citenamefont {Churaev},\ and\ \citenamefont
  {Muller}}]{surface_forces_1987}%
  \BibitemOpen
  \bibfield  {author} {\bibinfo {author} {\bibfnamefont {B.~V.}\ \bibnamefont
  {Derjaguin}}, \bibinfo {author} {\bibfnamefont {N.~V.}\ \bibnamefont
  {Churaev}}, \ and\ \bibinfo {author} {\bibfnamefont {V.~M.}\ \bibnamefont
  {Muller}},\ }\enquote {\bibinfo {title} {{Surface Forces in Transport
  Phenomena}},}\ in\ \href {\doibase 10.1007/978-1-4757-6639-4_11} {\emph
  {\bibinfo {booktitle} {{Surface Forces}}}}\ (\bibinfo  {publisher}
  {Springer},\ \bibinfo {address} {Boston},\ \bibinfo {year}
  {1987})\BibitemShut {NoStop}%
\bibitem [{\citenamefont {Derjaguin}\ and\ \citenamefont
  {Sidorenkov}(1941)}]{derjaguin_sidorenkov_1941thermoosmosis}%
  \BibitemOpen
  \bibfield  {author} {\bibinfo {author} {\bibfnamefont {B.~V.}\ \bibnamefont
  {Derjaguin}}\ and\ \bibinfo {author} {\bibfnamefont {G.~P.}\ \bibnamefont
  {Sidorenkov}},\ }\href@noop {} {\bibfield  {journal} {\bibinfo  {journal}
  {Doklady Akademii Nauk SSSR}\ }\textbf {\bibinfo {volume} {32}},\ \bibinfo
  {pages} {622} (\bibinfo {year} {1941})}\BibitemShut {NoStop}%
\bibitem [{\citenamefont {Piazza}\ and\ \citenamefont
  {Parola}(2008)}]{piazzaparola_2008}%
  \BibitemOpen
  \bibfield  {author} {\bibinfo {author} {\bibfnamefont {R.}~\bibnamefont
  {Piazza}}\ and\ \bibinfo {author} {\bibfnamefont {A.}~\bibnamefont
  {Parola}},\ }\href {http://stacks.iop.org/0953-8984/20/i=15/a=153102}
  {\bibfield  {journal} {\bibinfo  {journal} {Journal of Physics: Condensed
  Matter}\ }\textbf {\bibinfo {volume} {20}},\ \bibinfo {pages} {153102}
  (\bibinfo {year} {2008})}\BibitemShut {NoStop}%
\bibitem [{\citenamefont {Kennard}(1938)}]{kennard_1938kinetic}%
  \BibitemOpen
  \bibfield  {author} {\bibinfo {author} {\bibfnamefont {E.}~\bibnamefont
  {Kennard}},\ }\href@noop {} {\emph {\bibinfo {title} {Kinetic theory of
  gases: with an introduction to statistical mechanics}}}\ (\bibinfo
  {publisher} {McGraw-Hill},\ \bibinfo {year} {1938})\BibitemShut {NoStop}%
\bibitem [{\citenamefont {Maxwell}(1879)}]{maxwell_1879}%
  \BibitemOpen
  \bibfield  {author} {\bibinfo {author} {\bibfnamefont {J.~C.}\ \bibnamefont
  {Maxwell}},\ }\href {\doibase 10.1098/rstl.1879.0067} {\bibfield  {journal}
  {\bibinfo  {journal} {Philosophical Transactions of the Royal Society of
  London}\ }\textbf {\bibinfo {volume} {170}},\ \bibinfo {pages} {231}
  (\bibinfo {year} {1879})}\BibitemShut {NoStop}%
\end{thebibliography}%

\end{document}


\title{SUPPLEMENTARY MATERIAL - Thermal forces from a microscopic perspective}

\author{Pietro Anzini, Gaia Maria Colombo, Zeno Filiberti and Alberto Parola}
\affiliation{Dipartimento di Scienza e Alta Tecnologia, Universit\`a degli Studi dell'Insubria, Como, Italy}

\maketitle

This document provides some additional informations and the derivations of the main results of the letter {\it Thermal forces from a microscopic perspective}.

\tableofcontents

\makeatletter
\def\@ssect@ltx#1#2#3#4#5#6[#7]#8{%
  \def\H@svsec{\phantomsection}%
  \@tempskipa #5\relax
  \@ifdim{\@tempskipa>\z@}{%
    \begingroup
      \interlinepenalty \@M
      #6{%
       \@ifundefined{@hangfroms@#1}{\@hang@froms}{\csname @hangfroms@#1\endcsname}%
       {\hskip#3\relax\H@svsec}{#8}%
      }%
      \@@par
    \endgroup
    \@ifundefined{#1smark}{\@gobble}{\csname #1smark\endcsname}{#7}%
  }{%
    \def\@svsechd{%
      #6{%
       \@ifundefined{@runin@tos@#1}{\@runin@tos}{\csname @runin@tos@#1\endcsname}%
       {\hskip#3\relax\H@svsec}{#8}%
      }%
      \@ifundefined{#1smark}{\@gobble}{\csname #1smark\endcsname}{#7}%
      \addcontentsline{toc}{#1}{\protect\numberline{}#8}%
    }%
  }%
  \@xsect{#5}%
}%
\makeatother

\section{Microscopic conservation laws}
This Section provides the local expression of the fluxes which fulfil
the microscopic counterpart of the macroscopic conservation laws of mass, momentum and energy.
Furthermore the ambiguity in the definition of the fluxes arising in microscopic conservation laws is briefly discussed.

The general form of a local conservation law reads
\begin{equation}
\frac{\ud \hat A(\bmr)}{\ud t}  + \partial_\alpha \hat J_A^{\alpha}(\bmr) = 0,
\label{eq:cont_generale}
\end{equation}
where the ``operator'' $\hat A(\bmr)$ is the local conserved quantity, $\hat J_A^{\alpha}(\bmr)$ the corresponding current\footnote{Notice 
that here and in the following the dependence of the operators on the phase space variables is understood.} 
and $\partial_\alpha$ is the partial derivative w.r.t. $r^\alpha$.
The  mass, momentum and energy density operators have already been introduced in the main text; here we recall their definition
for future reference:
\begin{eqnarray}
\hat{\rho}(\bmr) &=& m\,\sum_i \delta(\bmq_i-\bmr),\label{eq:mass_dens} \\
\hat{j}^{\alpha}(\bmr) &=& \sum_i \delta(\bmq_i-\bmr)\, p_i^{\alpha} \vphantom{\sum_{j (\ne i)}\Bigg]},\label{eq:mom_dens} \\
\hat{\cal H}(\bmr) &=& \sum_i \delta(\bmq_i-\bmr)\,\hat{h}_i=
\sum_i \delta(\bmq_i-\bmr)\Bigg[ \frac{p_i^2}{2m} + \frac{1}{2} \sum_{j (\ne i)} v(|\bmq_i-\bmq_j|) +V(\bmq_i)\Bigg]. \label{eq:ham_dens}
\end{eqnarray}
In these expressions $\bmq_i$ and $\bmp_i$ are the generalised coordinates of the particle labelled by the index $i$, $m$ is the mass 
of the particles\footnote{Our approach can be straightforwardly generalised for particles with different masses.}, 
$v(|\bmq|)$ the inter-particle potential and $V(\bmq)$ the external potential which couples to the mass density.
We remark that, according to the {\it definition}~(\ref{eq:ham_dens}), 
the interaction energy $v_{ij}=v(|\bmq_i-\bmq_j|)$ between two particles $i$ and $j$ (located at $\bmq_i$ and $\bmq_j$)
is ascribed without justification half to particle $i$ and half to particle $j$. Another admissible definition 
of the local energy density could ascribe the whole interaction energy $v_{ij}$ to the point $(\bmq_i+\bmq_j)/2$.
The apparent ambiguity in~(\ref{eq:ham_dens}) is related to the non-local nature of the inter-particle 
interaction potential $v(\bmr)$~\cite{rowlinson_1993,irvingkirkwood_1949} and disappears when $\hat{\cal H}(\bmr)$ is 
integrated over the volume of the system (see also Ref.~\cite{baroni_15})
\begin{equation}
\hat{H}=\int \ud \bmr \, \hat{\cal H}(\bmr) = \sum_i\frac{p_i^2}{2m} + \frac{1}{2} \sum_{j \ne i} v(|\bmq_i-\bmq_j|) +\sum_i V(\bmq_i).
\label{eq:ham}
\end{equation}

According to Eq.~(\ref{eq:cont_generale}) the formal expression of the fluxes directly follows from the evaluation of the 
time derivative of the conserved density, which reduces to the action of the Liouvillian ${\scr L}$ 
(corresponding to the Hamiltonian~(\ref{eq:ham})) on the density $\hat A(\bmr)$:
\begin{equation}
\frac{\ud \hat A(\bmr)}{\ud t} = - {\scr L} \hat A(\bmr).
\label{eq:evo_liouvillian}
\end{equation}
The expression of ${\scr L}$ is reported in standard textbooks (see e.g. Ref.~\cite{balescu_equilibrium_1975}) and 
for the Hamiltonian~(\ref{eq:ham}) reads  
\begin{eqnarray}
{\scr L}&=&{\scr L}_K+{\scr L}_v+{\scr L}_V \notag \\
&=&- \sum_i\frac{\bmp_i}{m}\cdot \frac{\partial}{\partial \bmq_i}
+\frac{1}{2} \sum_{i\ne j}\frac{\partial v_{ij}}{\partial \bmq_i}\cdot\left(\frac{\partial}{\partial \bmp_i}-\frac{\partial}{\partial \bmp_j}\right)
+\sum_i\frac{\partial V(\bmq_i)}{\partial \bmq_i}\cdot\frac{\partial}{\partial \bmp_i}.
\label{eq:liouvillian_centraladditive}
\end{eqnarray}

\subsection*{Density conservation}

The continuity equation for the mass density~(\ref{eq:mass_dens}) directly provides the corresponding mass current: 
\begin{align}
\frac{\ud \hat{\rho}(\bmr)}{\ud t} &=-{\scr L} \, \hat{\rho}(\bmr)=\sum_i \frac{\partial}{\partial \bmq_i} \delta(\bmq_i-\bmr) \cdot \bmp_i \notag\\
&= - \partial_{\alpha} \, \hat{j}^{\alpha}_\rho(\bmr) 
\label{eq:masscons}
\end{align}
from which\footnote{Notice the notation adopted here is in accordance with Eq.~(\ref{eq:cont_generale}). This choice is 
justified by the fact that the mass current $\hat{j}^{\alpha}_\rho(\bmr)$ resulting from Eq.~(\ref{eq:masscons}) is defined up to a divergence-free scalar field. 
On the other hand, the operator $\hat{j}^{\alpha}(\bmr)$, related to the observable momentum density, has been defined unambiguously in Eq.~(\ref{eq:mom_dens}).}
\begin{equation}
 \hat{j}^{\alpha}_\rho(\bmr)=\sum_i \delta(\bmq_i-\bmr)p_i^{\alpha}.
\notag
\end{equation}

\subsection*{Momentum conservation}

Analogously, the local conservation law corresponding to the macroscopic momentum balance equation can be obtained evaluating  
the rate of change of the momentum density $\hat{j}^{\alpha}(\bmr)$ in Eq.~(\ref{eq:mom_dens}):
\begin{equation}
\frac{\ud \hat{j}^{\alpha}(\bmr)}{\ud t}=-{\scr L} \, \hat{j}^{\alpha}(\bmr)
=-\partial_{\beta}\left[\sum_i \frac{p_i^{\alpha}p_i^{\beta}}{m} \delta(\bmq_i-\bmr)\right] 
-\frac{\hat{\rho}(\bmr)}{m}{\partial_\alpha V(\bmr)}
-\frac{1}{2}\sum_{i\neq l}\frac{\partial v_{il}}{\partial q_i^{\alpha}}\bigl[\delta(\bmq_i-\bmr)-\delta(\bmq_l-\bmr)\bigr].
\label{eq:momcons_first}
\end{equation}
The last term in Eq.~(\ref{eq:momcons_first}) can be written as the divergence of a second rank tensor 
by means of the distributional identity~\cite{schofield_henderson_1982}\footnote{As shown by Irving and Kirkwood~\cite{irvingkirkwood_1949}, 
an analogous result follows from a {\it formal} Taylor's series expansion of $\delta(\bmq_j-\bmr)-\delta(\bmq_i-\bmr)$
in the vector separation $\bmq_j-\bmq_i$.} 
\begin{equation}
\delta(\bmq_j-\bmr)-\delta(\bmq_i-\bmr)=
\oint_{C_{i\rightarrow j}} \ud y^{\gamma} \frac{\partial}{\partial y^{\gamma}}\delta\left(\bm{y}-\bmr\right)=
-\partial_{\gamma}\oint_{C_{i\rightarrow j}} \ud y^{\gamma}\delta\left(\bm{y}-\bmr\right),
\label{eq:id_deltas}
\end{equation}
where the integral is along {\it any} contour $C_{i\rightarrow j}$ from $\bmq_i$ to $\bmq_j$.
Making use of this result in~(\ref{eq:momcons_first}) we obtain the microscopic continuity equation 
for the momentum density $\hat{j}^{\alpha}(\bmr)$
\begin{equation}
\frac{\ud \hat{j}^{\alpha}(\bmr)}{\ud t}=-\partial_\beta \hat{J}_j^{\alpha\beta}(\bmr)-\frac{\hat{\rho}(\bmr)}{m}{\partial_\alpha V(\bmr)},
\label{eq:momcons}
\end{equation}
where the microscopic momentum current operator $\hat{J}_j^{\alpha\beta}(\bmr)$ has been defined as\footnote{Some references adopt a 
slightly different notation, introducing the {\it stress tensor} $\hat{\sigma}^{\alpha\beta}$  
defined as $\hat{\sigma}^{\alpha\beta}(\bmr)=-\hat{J}_j^{\alpha\beta}(\bmr)$.} 
\begin{eqnarray}
\hat{J}_j^{\alpha\beta}(\bmr)&=&\sum_i \frac{p_i^{\alpha}p_i^{\beta}}{m} \delta(\bmq_i-\bmr)+
\frac{1}{2}\sum_{i\neq l}\frac{\partial v_{il}}{\partial q_i^{\alpha}}\oint_{C_{i\rightarrow l}} \ud y^{\beta}\delta\left(\bm{y}-\bmr\right)
\label{eq:mom_flow}\\
&=&\sum_i \frac{p_i^{\alpha}p_i^{\beta}}{m} \delta(\bmq_i-\bmr)-
\frac{1}{2}\sum_{i\neq l}\frac{q_{il}^{\alpha}}{|\bmq_{il}|}\left.\frac{\ud v(q)}{\ud q}\right|_{q=|\bmq_{il}|}\oint_{C_{i\rightarrow l}} \ud y^{\beta}\delta\left(\bm{y}-\bmr\right),
\label{eq:mom_flow1}
\end{eqnarray}
where~(\ref{eq:mom_flow1}) holds when the particles interacting through a central potential and $\bmq_{ij}=\bmq_j-\bmq_i$.
Notice that the last term in Eq.~(\ref{eq:momcons}) acts as a source contribution when a space-dependent external field $V(\bmr)$ is present.
The average value of the operator $\hat{J}_j^{\alpha\beta}(\bmr)$ is the so called (local) {\it pressure tensor}
\begin{equation}
p^{\alpha \beta}(\bmr)=\big< \hat{J}_j^{\alpha\beta}(\bmr)\big>.
\label{eq:def_presstens}
\end{equation}
Equation~(\ref{eq:mom_flow}) shows how the local momentum current, which enters the continuity equation for the momentum density~(\ref{eq:momcons}), 
can not be defined without ambiguity: Different contours in Eq.~(\ref{eq:mom_flow}) lead to different expressions for $\hat{J}_j^{\alpha\beta}(\bmr)$,
and the same considerations also apply to the pressure tensor~(\ref{eq:def_presstens}). 
The average of Eq.~(\ref{eq:mom_flow1}) provides the local pressure tensor in a system where particles interact through a central \hbox{pair-wise} 
additive potentials\footnote{See~\cite{schofield_henderson_1982}, Eq. (3.2).}:
\begin{equation}
p^{\alpha \beta}(\bmr)=\frac{\rho(\bmr)\,k_{\mathrm{B}}T}{m} \, \delta^{\alpha\beta}-
\frac{1}{2}\int \ud \bmy \, \frac{y^\alpha}{|\bmy|}\frac{\ud v(|\bmy|)}{\ud |\bmy|} 
\oint_{C_{{\bm 0}\rightarrow\bmy}} \ud s^{\beta} \rho^{(2)}(\bmr-\bms,\bmr-\bms+\bmy).
\label{eq:ppairwise}
\end{equation}
Here $\rho^{(2)}(\bmr,\bmr')$ is the \hbox{two-particle} density \cite{simpleliquids_fourth} and the \hbox{line-integral} is extended, without
any loss in generality~\cite{schofield_henderson_1982}, from the origin $\bm 0$ to a given point $\bmy$. 
Making use of this result it is possible to show that in the homogeneous and isotropic limit the ambiguity in the 
definition of the pressure tensor disappears. Indeed the two-particle distribution function can be expressed in terms of
the radial distribution function 
\begin{equation}
\rho^{(2)}(\bmr,\bmr')=\frac{\rho^2}{m^2}\, g(|\bmr-\bmr'|) 
\notag
\end{equation}
and finally Eq.~(\ref{eq:ppairwise}) reduces to
\begin{equation}
p^{\alpha \beta}(\bmr)=p \, \delta^{\alpha \beta}=\frac{\rho \,k_{\mathrm{B}}T}{m} \, \delta^{\alpha\beta} -
\frac{1}{2}\frac{\rho^2}{m^2}\int \ud \bmr \, \frac{r^\alpha r^\beta}{|\bmr|}\frac{\ud v(|\bmr|)}{\ud |\bmr|} \, g(|\bmr|)
\label{eq:viriale}
\end{equation}
which is the well known {\it virial} (or pressure) equation for a homogeneous and isotropic fluid 
at density $\rho$~\cite{simpleliquids_fourth}.

\subsubsection*{Comment on the ambiguity in the definitions of the fluxes}

The non-uniqueness of the local pressure tensor has been implicitly recognised by Kirkwood 
in the fifties. He obtained an expression for the configurational contribution to the stress tensor 
in a paper with Buff~\cite{kirkwoodbuff_1949} and a different one in another work with Irving~\cite{irvingkirkwood_1949}\footnote{The 
so called \hbox{Irving-Kirkwood} stress tensor is reported in the Appendix of~\cite{irvingkirkwood_1949}.}. 
Harasima gave in 1958 the first explicit description of this ambiguity~\cite{harasima_1958} and a review was published 
by Ono and Kondo~\cite{ono_kondo_60} a couple of years later. 
A rigorous and exhaustive study of the problem was given in the eighties by Schofield and Henderson~\cite{schofield_henderson_1982}.
More recently, Baus and Lovett~\cite{baus_lovettPRL90,baus_lovettPRA91} (and also other authors) have attempted to define the pressure 
tensor uniquely. However, their definition can not be accepted because it only holds for particular geometries~\cite{rowlinson_replyPRL,rowlinson_1993}.

This ambiguity related to the definition of the pressure tensor has been recovered in two papers~\cite{ganti_2017,ganti_2018} 
published a couple of years ago dealing with \hbox{thermo-osmosis}. The authors try to discriminate between different expressions of the 
pressure tensor estimating the value of the \hbox{thermo-osmotic} flow resulting from (approximate) predictions 
which involve the knowledge of the pressure tensor itself. 
In the most recent paper~\cite{ganti_2018}, they compare these predictions with the 
(exact) results obtained through a clever nonequilibrium molecular dynamics simulation and they conclude that both the virial and 
the \hbox{Irving-Kirkwood} expression do not accurately predict surface forces due to temperature gradients.
\newline
However, we remark that the infinite possible definitions of the the pressure tensor are indeed equivalent, i.e. 
all the {\it physical observables} must be invariant with respect to different choices of the path $C_{ij}$~\cite{schofield_henderson_1982}. 
As regards an inhomogeneous fluid, the pressure tensor itself is {\it not} a well defined observable on a length scale shorter than the correlation length 
or the range of the \hbox{inter-particle} potential~\cite{rowlinson_1993}. Qualitatively, we can try to understand this circumstance
reflecting on the fact that it is not possible to identify the surface where the pressure is acting. Analogously,
we can not define without ambiguities the surface which separate two different phases of the same fluid. On the other 
hand, the pressure exerted on a given region of fluid and the surface tension of an interface are both well defined observables 
and indeed it is possible to prove that they do not depend on the particular definition of the pressure tensor~\cite{schofield_henderson_1982}\footnote{The virial expression is an allowed choice for the pressure tensor 
only for homogeneous fluids. See below.}.
\newline
As regards {\it approximate theories}, such as the local equilibrium assumption or the approach originally put forward by Derjaguin 
and recently \hbox{re-derived} in~\cite{ganti_2017,ganti_2018}, the invariance of the observables with respect different definitions of 
the pressure tensor is not guaranteed a priori. However, the slip velocity of a fluid subject to a temperature gradient is 
a genuine physical quantity, also from the microscopic viewpoint. 
Therefore every {\it exact} prediction of the \hbox{thermo-osmotic} slip must be invariant 
on the choice of the trajectory in~(\ref{eq:mom_flow}): We conclude that both the local thermal equilibrium and 
the Derjaguin approach should be considered as approximations, because their expression are not endowed by this invariance. 
\newline
Finally, we remark that the virial pressure tensor~(\ref{eq:viriale}) extended to inhomogeneous systems, 
which has been evaluated in~\cite{ganti_2017,ganti_2018} in order to obtain the \hbox{thermo-osmotic} properties of the fluid, 
does not correspond to any choice of the path in~(\ref{eq:mom_flow}) and in addition it does not fulfil the hydrostatic balance condition. 
This expression is commonly adopted within continuum hydrodynamics, where it is assumed that the relevant quantities
vary on a length scale much larger than the correlation length\footnote{Unless the pathological condition where the system is in near critical conditions.}.

\subsection*{Energy conservation}
The microscopic conservation law for the energy density $\hat{\cal H}(\bmr)$ can be obtained through
the same steps followed before in the case of the mass and momentum current:
\begin{equation}
\frac{\ud \hat{\cal H}(\bmr)}{\ud t} =-{\scr L} \, \hat{\cal H}(\bmr)
=-{\scr L}_K\hat{\cal H}(\bmr)-\sum_i \delta(\bmq_i-\bmr)\Bigl[{\scr L}_v+{\scr L}_V\Bigr] \frac{p_i^2}{2m}. 
\label{eq:contH_prima}
\end{equation}
After some algebra, the action of the Liouvillians on the Hamiltonian reads
\begin{eqnarray}
{\scr L}_K\hat{\cal H}(\bmr)&=&\sum_i\frac{p^{\alpha}_i}{m}\left[h_i \partial_\alpha\delta(\bmq_i-\bmr)
-\frac{\partial V(\bmq_i)}{\partial q_i^{\alpha}} \right]
-\frac{1}{2m}\sum_{i\neq j} \delta(\bmq_i-\bmr)\frac{\partial v_{ij}}{\partial q_i^{\alpha}}\left(p_i^{\alpha}-p_j^{\alpha}\right),\notag\\
\Bigl[{\scr L}_v+{\scr L}_V\Bigr]\frac{p_i^2}{2m}&=&\frac{p^{\alpha}_i}{m}\left[\sum_{j(\neq i)}\frac{\partial v_{ij}}{\partial q_i^{\alpha}}
+\frac{\partial V(\bmq_i)}{\partial q_i^{\alpha}}\right].\notag
\end{eqnarray}
Making use of these results in Eq.~(\ref{eq:contH_prima}), we get
\begin{equation}
\frac{\ud \hat{\cal H}(\bmr)}{\ud t} =-\partial_\alpha \left[\sum_i\delta(\bmq_i-\bmr)\frac{p^{\alpha}_i}{m}\hat{h}_i\right]
- \frac{1}{2m}\sum_{i\neq j}p_i^{\alpha}\frac{\partial v_{ij}}{\partial q_i^{\alpha}}\big[ \delta(\bmq_i-\bmr)-\delta(\bmq_j-\bmr)\big].  
\label{eq:contHseconda}
\end{equation}
The identity~(\ref{eq:id_deltas}) allows to write Eq.~(\ref{eq:contHseconda}) in the form of a microscopic 
conservation law
\begin{equation}
\frac{\ud \hat{\cal H}(\bmr)}{\ud t} = -\partial_\alpha \hat{J}_{\cal H}^\alpha(\bmr), 
\label{eq:hamcons}
\end{equation}
where we have defined the energy current $\hat{J}_{\cal H}^\alpha(\bmr)$ as
\begin{equation}
\hat{J}_{\cal H}^\alpha(\bmr)=\sum_i\frac{p^{\alpha}_i}{m}\delta(\bmq_i-\bmr)\hat{h}_i+
\frac{1}{2}\sum_{i}\frac{p^{\delta}_i}{m}\sum_{j(\neq i)}\frac{\partial v_{ij}}{\partial q_i^{\delta}}\oint_{C_{i\rightarrow j}} \ud y^{\alpha}\delta\left(\bm{y}-\bmr\right).
\label{eq:form_def_JH}
\end{equation}
Here we stress that~(\ref{eq:form_def_JH}) is the microscopic energy flux according to the definition of 
the local energy density given in Eq.~(\ref{eq:ham_dens}). Different microscopic forms of the local energy
provide different expressions of $\hat{J}_{\cal H}^\alpha(\bmr)$. In addition to this, the same considerations 
stated above for the momentum current apply: The ambiguity in the definition of the heat flux
is recovered in the freedom connected to the choice of the path.
However the thermal transport coefficients, which are genuine physical observables, turn out to be independent 
on the particular choice in Eq.s~(\ref{eq:form_def_JH}) and~(\ref{eq:ham_dens}) (see also Ref.~\cite{baroni_15}).

\section{Evaluation of averages}

In this Section we evaluate the averages according to the local equilibrium distribution
\begin{equation}
F^{{LE}} = \mathcal{Q}^{-1} \,\mathrm{e}^{-\int \ud\bmr \,\beta(\bmr)\, \hat{\cal E}(\bmr)},
\label{eq:fle}
\end{equation}
and the dynamic corrections to $F^{{LE}}$ induced by ensuing dynamics (see the main text).
\newline
In Eq.~(\ref{eq:fle}) $\beta(\bmr)$ is a scalar field related to the (inverse) local temperature, 
\begin{equation}
\mathcal{Q}= \mathrm{Tr}\left\{\mathrm{e}^{-\int \ud\bmr \,\beta(\bmr)\, \hat{\cal E}(\bmr)}\right\}
\notag
\end{equation}
is the partition function\footnote{The symbol $\mathrm{Tr}\{\dots\}$ is the abbreviation of the trace over all the degrees of freedom.} 
and the local internal energy operator~$\hat{\cal E}(\bmr)$, already defined in the main text, reads 
\begin{equation}
\hat{\cal E}(\bmr) = \hat{\cal H}(\bmr)-\hat{\bmj}(\bmr)\cdot \bmu(\bmr)-\mu(\bmr)\hat{\rho}(\bmr).
\notag
\end{equation}
The local Hamiltonian, momentum and particle densities have been defined in Eq.~(\ref{eq:ham_dens}),~(\ref{eq:mom_dens}) 
and~(\ref{eq:mass_dens}) respectively; $\bmu(\bmr)$ and $\mu(\bmr)$ are 
the vector and scalar fields related to the local velocity profile and chemical potential (per unit mass) 
of the fluid respectively. 

\subsection{Local Equilibrium averages}

The averages according to the static LE distribution~(\ref{eq:fle}) can be evaluated within linear response theory 
as follows. The essential hypothesis is that the nonequilibrium state defined by~(\ref{eq:fle}) is very close,
or analogously a small perturbation, of an equilibrium state.
First of all we introduce the underlying \hbox{(zero-order)} equilibrium distribution function
\begin{equation}
F^{eq}=\mathcal{Q}_0^{-1}\ue^{-\beta\left(H-m  \mu  N\right)},
\label{eq:fequnder}
\end{equation}
defined by the constant average temperature $\beta$ and average chemical potential (per unit mass) $\mu$ and characterised by
a vanishing velocity field $\bmu(\bmr)=0$. Here $\mathcal{Q}_0$ is the equilibrium grand canonical partition function. 
The fields characterising the \hbox{out-of-equilibrium} state can then be written in terms of small deviations 
from the constant values of the temperature, the chemical potential and the vanishing velocity:
\begin{equation}
\beta(\bmr)=\beta+\delta\beta(\bmr), \qquad \quad \mu(\bmr)=\mu+\delta\mu(\bmr), \qquad \quad \bmu(\bmr)={\bm 0}+\delta\bmu(\bmr).
\notag
\end{equation}
Following the method inspired by linear response theory, we expand the LE distribution~(\ref{eq:fle}) about 
the equilibrium distribution~(\ref{eq:fequnder}) to the first order in the deviations $\delta\beta(\bmr)$,
$\delta\mu(\bmr)$ and $\delta\bmu(\bmr)$. Noticing that in~(\ref{eq:fle}) the deviations from the underlying equilibrium distribution
arise both in the exponential and in the partition function, we obtain
\begin{equation}
F^{LE}=\frac{\ue^{-\int \ud\bmr \,\beta(\bmr)\, \hat{\cal E}(\bmr)}}{\mathcal{Q}}\simeq\frac{\mathcal{Q}_0\,F^{eq}\,(1-\hat{C}_e)}{\mathcal{Q}_0\,(1-C_\mathcal{Q})}\simeq F^{eq}(1-\hat{C}_e+C_\mathcal{Q}),
\notag
\end{equation}
where the linear corrections to the exponential and the partition function can be written as
\begin{eqnarray}
\hat{C}_e&=&\int \ud \bmr \biggl\{\delta\beta(\bmr)\left[\hat{\cal H}(\bmr)-\mu \hat{\rho}(\bmr)\right]
-\beta\left[\hat{\bmj}(\bmr)\cdot \delta\bmu(\bmr)+\delta\mu(\bmr)\hat{\rho}(\bmr)\right] \biggr\}, \notag \\
C_\mathcal{Q}&=&\int \ud \bmr\, \biggl\{\delta\beta(\bmr) \left[\big< \hat{\cal H}(\bmr)\big>_0 -\mu \big<\hat{\rho}(\bmr)\big>_0\right]
+ \beta\delta\mu(\bmr)\big< \hat{\rho}(\bmr)\big>_0	\biggr\}.
\label{eq:c_q_localeq}
\end{eqnarray}
The averages $\left<\dots\right>_0$ are evaluated according to the equilibrium distribution~(\ref{eq:fequnder})
and the different notation between $\hat{C}_e$ and $C_\mathcal{Q}$ underlines that $\hat{C}_e$ still depends on the phase-space coordinates.
The final expression for the LE distribution within the linear approximation is given by
\begin{equation}
F^{LE}=F^{eq}\Biggl\{1-\int \ud \bmr \biggl\{\delta\beta(\bmr)\left[\hat{\cal H}(\bmr)-\mu \hat{\rho}(\bmr)\right]
-\beta\left[\hat{\bmj}(\bmr)\cdot \delta\bmu(\bmr)+\delta\mu(\bmr)\hat{\rho}(\bmr)\right] \biggr\}+ C_\mathcal{Q} \Biggr\}
\label{eq:fle_linear}
\end{equation}
and the local equilibrium average of a given observable $\hat{A}(\bmr)$ reads
\begin{align}
\big<\hat{A}(\bmr)\big>_{LE}&=\big<\hat{A}(\bmr)\big>_0-\int \ud \bmr' \biggl\{\delta\beta(\bmr')
\Bigl[\big<\hat{A}(\bmr)\,\hat{\cal H}(\bmr')\big>_0-\mu \big<\hat{A}(\bmr)\,\hat{\rho}(\bmr')\big>_0\Bigr] \notag \\
& \qquad \qquad \qquad -\beta\Bigl[\big<\hat{A}(\bmr)\,\hat{j}^{\alpha}(\bmr')\big>_0 \delta u^{\alpha}(\bmr')
+\delta\mu(\bmr')\big<\hat{A}(\bmr)\hat{\rho}(\bmr')\big>_0\Bigr] \biggr\}+ C_\mathcal{Q}\big<\hat{A}(\bmr)\big>_0.
\label{eq:fle_average}
\end{align}

This result can be immediately applied to the relevant observables specified by the momentum density, the energy current and the particle density operators:
\begin{eqnarray}
\big< \hat{j}^{\alpha}(\bmr)\big>_{LE}&=&\beta \int \ud\bmr^\prime \big<\hat{j}^{\alpha}(\bmr)\hat{j}^{\gamma}(\bmr')\big>_0 u^{\gamma}(\bmr')
=\left< \hat{\rho}(\bmr)\right>_0u^{\alpha}(\bmr),\notag\\
\big< \hat{J}_{\cal H}^{\alpha}(\bmr)\big>_{LE}&=&
\beta \int \ud\bmr^\prime \big<\hat{J}_{\cal H}^{\alpha}(\bmr)\,\hat{j}^{\gamma}(\bmr')\big>_0\, u^{\gamma}(\bmr'), \notag\\
\left< \hat{\rho}(\bmr)\right>_{LE}&=&\left<\hat{\rho}(\bmr)\right>_{0}
-\int \ud \bmr' \biggl\{\delta\beta(\bmr')\Bigl[\big<\hat{\rho}(\bmr)\,\hat{\cal H}(\bmr')\big>_0
-\mu \left<\hat{\rho}(\bmr)\,\hat{\rho}(\bmr')\right>_0\Bigr]
-\beta\, \delta\mu(\bmr')\left<\hat{\rho}(\bmr)\,\hat{\rho}(\bmr')\right>_0 \biggr\}
+ C_\mathcal{Q}\left<\hat{\rho}(\bmr)\right>_0\notag\\
&=&\left<\hat{\rho}(\bmr)\right>_{0}\Big|_{\beta(\bmr),\mu(\bmr)},
\label{eq:rho_le_generale}
\end{eqnarray}
where the last identity~(\ref{eq:rho_le_generale}) shows that the local equilibrium average of the density can be evaluated 
according to the equilibrium average~(\ref{eq:fequnder}), with the temperature and the 
chemical potential fixed at their local value in $\bmr$, that is $\beta(\bmr)$ and $\mu(\bmr)$.
\newline
The local equilibrium average of the momentum current deserves special mention:
\begin{align}
\big< \hat{J}_{j}^{\alpha\beta}(\bmr)\big>_{LE}&=\big<\hat{J}_{j}^{\alpha\beta}(\bmr)\big>_{0}
-\int \ud \bmr' \biggl\{\delta\beta(\bmr')\Bigl[\big<\hat{J}_{j}^{\alpha\beta}(\bmr)\,\hat{\cal H}(\bmr')\big>_0
-\mu \big<\hat{J}_{j}^{\alpha\beta}(\bmr)\,\hat{\rho}(\bmr')\big>_0\Bigr] \notag \\
& \qquad \qquad \qquad \qquad \qquad \quad \qquad \qquad -\beta\, \delta\mu(\bmr')\big<\hat{J}_{j}^{\alpha\beta}(\bmr)\,\hat{\rho}(\bmr')\big>_0 \biggr\}
+ C_\mathcal{Q}\big<\hat{J}_{j}^{\alpha\beta}(\bmr)\big>_0.
\label{eq:Jj_le_generale}
\end{align}
Indeed, this local equilibrium average is characterised by \hbox{non-zero} \hbox{off-diagonal} elements 
because, due to the configurational contribution in~(\ref{eq:mom_flow}),
$\hat{J}_{j}^{\alpha\beta}(\bmr)$ is not an odd operator with respect to the momenta.
It follows that the momentum flux tensor, which is diagonal in equilibrium systems, can acquire \hbox{off-diagonal} components 
when the state of the system is described by a LE distribution as (\ref{eq:fle}). 
However, it can be useful to point out that the {\it diagonal} components of this tensor can be written making use of the shorthand notation
introduced in Eq.~(\ref{eq:rho_le_generale}) as
\begin{equation}
\big< \hat{J}_{j}^{\alpha\alpha}(\bmr)\big>_{LE}=
\big< \hat{J}_{j}^{\alpha\alpha}(\bmr)\big>_{0}\Big|_{\beta(\bmr),\mu(\bmr)}=p^{\alpha\alpha}(\bmr)\Big|_{\beta(\bmr),\mu(\bmr)},
\label{eq:jj_trick}
\end{equation}
where $p^{\alpha\beta}(\bmr)$ is the equilibrium pressure tensor.
\newline
Notice that, if the local equilibrium state is defined through the correct mathematical expression~(\ref{eq:fle}), 
the simple trick introduced in \hbox{Eq.s~(\ref{eq:rho_le_generale})} \hbox{and~(\ref{eq:jj_trick})} and employed in many references 
for the evaluation of the LE averages, is not valid for all the observables.

\subsection{Dynamic corrections to the Local Equilibrium averages}
The LE distribution function~(\ref{eq:fle}) introduced above is not stationary under the action of the Liouvillian  
\hbox{(${\scr L} F^{LE}\ne 0$)} and it can not be used to evaluate averages in stationary 
conditions. Indeed, if the external constraints\footnote{I.e. the temperature, the chemical potential and the velocity fields.} 
are kept fixed, the actual \hbox{phase-space} distribution 
will evolve in time due to the ensuing dynamics towards a stationary \hbox{(time-independent)} \hbox{out-of-equi}librium distribution.
In order to obtain such distribution we follow, with slight changes, the approach proposed by Mori~\cite{mori_1956,mori_1958}.  
Let us {\it assume} that the system is described at $t=0$ by a given LE state $F(t=0)=F^{LE}$. At times $t>0$ the \hbox{phase-space} 
distribution $F(t)$ will evolve according to the Liouville equation
\begin{equation}
\partial_t F(t) = {\scr L}F(t).
\label{eq:liouville} 
\end{equation}
The formal solution of~(\ref{eq:liouville}) can be written in integral form as
\begin{eqnarray}
F(t)&=&F(0) + \int_0^t \ud t^\prime \, \frac{\ud}{\ud t^\prime} F(t^\prime) \notag \\
&=&F^{LE} + \int_0^t \ud t^\prime \, {\scr L} \, {\scr U}(t^\prime)F^{LE} \notag\\ 
&=&F^{LE}+\int_0^t \ud t^\prime \,{\scr U}(t^\prime) \Bigl[{\scr L}F^{LE}\Bigr], \label{eq:f_toperatorU}
\end{eqnarray}
where ${\scr U}(t)=\mathrm{exp}\{t{\scr L}\}$ is the time evolution operator associated to the Liouvillian ${\scr L}$.
\newline
The explicit evaluation of the right hand side of~(\ref{eq:f_toperatorU}), to linear order in the field $\bmu(\bmr)$ and in the gradients 
$\partial_\alpha \beta(\bmr)$, $\partial_\alpha \mu(\bmr)$ and $\partial_\alpha u^{\beta}(\bmr)$, is straightforward and reads
\begin{equation}
F(t)=F^{LE}-\int_0^t \ud t^\prime \int \ud \bmr \,{\scr U}(t^\prime)\biggl\{F^{LE}\,\beta(\bmr)\,
\Bigl[ \partial_\alpha \hat{J}_{\cal H}^\alpha(\bmr)-\partial_\gamma \hat{J}_j^{\alpha\gamma}(\bmr) u^{\alpha}(\bmr)
-\mu(\bmr) \partial_{\alpha} \, \hat{j}_{\rho}^{\alpha}(\bmr)\Bigr]\biggr\}
\label{eq:dens_prob_Ftermini}
\end{equation}
where the action of the Liouvillian on the collisional invariants has been evaluated in the previous Section (see \hbox{Eq.s~(\ref{eq:hamcons})},~(\ref{eq:momcons}) 
and~(\ref{eq:masscons})).
\newline
Assuming that the perturbation on the system due to the fields $\beta(\bmr)$, $\bmu(\bmr)$ and $\mu(\bmr)$ 
is small we can restrict to a linear response approach and the \hbox{(time-dependent)} average of a given local observable $\hat{A}(\bmr)$ reads
\begin{equation}
\big< \hat{A}(\bmr) \big>_t=\big< \hat{A}(\bmr) \big>_{LE}-
\int_0^t \ud t^\prime \int \ud \bmr' \,\mathrm{Tr}\Biggl\{\hat{A}(\bmr)\,{\scr U}(t^\prime)\biggl[F^{LE}\,\beta(\bmr')
\Bigl(\dep_\alpha \hat{J}_{\cal H}^\alpha(\bmr')-\dep_\gamma \hat{J}_j^{\alpha\gamma}(\bmr') u^{\alpha}(\bmr')
-\mu(\bmr')\dep_{\alpha} \, \hat{j}_{\rho}^{\alpha}(\bmr')\Bigr)\biggr]\Biggr\},
\notag 
\end{equation}
where $\dep_\alpha$ is the derivative w.r.t. ${r'}^\alpha$.
Integrating by parts and taking the limit $t \to \infty$ we obtain
\begin{align}
\big< \hat{A}(\bmr) \big>=\big<\hat{A}(\bmr) \big>_{LE}+
\int_0^\infty \ud t^\prime \int \ud \bmr' \,
\biggl[ \big< \hat{A}(\bmr,t')\, \hat{J}_{\cal H}^\alpha(\bmr')\big>_0 \, \dep_\alpha\beta(\bmr')
- &\beta \big< \hat{A}(\bmr,t') \,\hat{J}_j^{\alpha\gamma}(\bmr')\big>_0 \, \dep_\gamma u^{\alpha}(\bmr') \notag \\
& \qquad -  \big<\hat{A}(\bmr,t') \, \hat{j_{\rho}}^{\alpha}(\bmr')\big>_0 \,\dep_{\alpha}\bigl[\beta\mu\bigr](\bmr')\biggr], 
\label{eq:aver_mori_gen}
\end{align}
where have shifted the time dependence on the observable $\hat{A}(\bmr)$ performing the canonical 
transformation ${\scr U}(-t')$ and we have retained only the {\it linear} contributions as in Eq.~(\ref{eq:dens_prob_Ftermini}).

\section{Constraints for the external fields}

Equation~(\ref{eq:aver_mori_gen}), together with the LE average~(\ref{eq:fle_average}), allows to evaluate the average of the relevant observables
for an \hbox{out-of-equilibrium} system. However, the \hbox{time-independent} fields $\beta(\bmr)$,
$\mu(\bmr)$ and $\bmu(\bmr)$, which enter these expressions, have not been fixed yet. Actually they can not be determined {\it a priori}.
In order to obtain their values, we introduce some additional informations about the system.

\subsection{Stationary continuity equations}

We begin by imposing the physical constraints characterising a stationary state, 
namely the {\it stationary} continuity equations satisfied by the average local energy density
$\big<\hat{\cal H}(\bmr)\big>$, by the average local momentum density $\big<\hat{j}^\gamma(\bmr)\big>$ 
and by the average local particle density $\left<\hat{\rho}(\bmr)\right>$. In formulae:
\begin{eqnarray}
\partial_\gamma\,\big<\hat{j}^\gamma(\bmr)\big> &=& 0, \label{eq:stat1} \\
\partial_\gamma\,\big<\hat{J}_{\cal H}^\gamma(\bmr)\big> &=&0, \vphantom{\frac{\left<\hat{\rho}(\bmr)\right>}{m}} \label{eq:stat2} \\
\partial_\gamma\,\big<\hat{J}_j^{\alpha\gamma}(\bmr)\big> &=&  -
\frac{\left<\hat{\rho}(\bmr)\right>}{m}\,\partial_\alpha V(\bmr). \label{eq:stat3}
\end{eqnarray}
The solution of this set of five independent differential equations formally provides the gradients of the fields 
$\partial_\alpha\beta(\bmr)$, $\partial_\alpha\mu(\bmr)$ and $\partial_\alpha u^{\gamma}(\bmr)$.
Unfortunately, without further approximations the general solution of this system can not be obtained in closed form. 
On the other hand, when the equations are specialised to some simple geometry, 
symmetry considerations allow to considerably simplify the problem. 

\begin{figure}
\centering
\begin{tikzpicture}[scale=0.8]

\shade[top color=slab,bottom color =white] (0,-0.5) rectangle (12,0.8) ;
\shade[top color=white,bottom color =slab] (0,4.3) rectangle (12,3) ;
\shade[top color=liquid,bottom color =liquid] (0,3) rectangle (12,0.8) ;

\draw[thick,->] (0.8,0.8) -- (11.2,0.8) node[below right,xshift=-0.45cm, yshift=-1pt] {$x$};
\draw[thick,->] (6,0.2) -- (6,3.9) node[below right,xshift=-0.5cm, yshift=0pt] {$z$};

\draw (6,0.8) node[anchor=east,yshift=-7pt] {$0$};
\draw (6,3) node[anchor=east,xshift=-2pt,yshift=1pt] {$h$};
\draw (6,1.9) node[anchor=east,xshift=-2pt,yshift=1pt] {$h/2$};
\draw[thick] (6cm+2pt,3cm) -- (6cm-2pt,3cm) node[anchor=east] {};
\draw[thick] (6cm+2pt,1.9cm) -- (6cm-2pt,1.9cm) node[anchor=east] {};

\draw[thick,->,arrows={-latex},grad] (7.5,1.9) -- (8.5,1.9) node[above,xshift=-0.5cm, yshift=0pt] {};
\draw[thick] (8,1.9) node[above,xshift=0cm, yshift=0pt] {$\partial_x \beta$};

\end{tikzpicture}
\caption{Schematic representation of the slab geometry. The $y$ direction is perpendicular to the plane of the sheet.}
\label{fig:geoslab}

\end{figure}
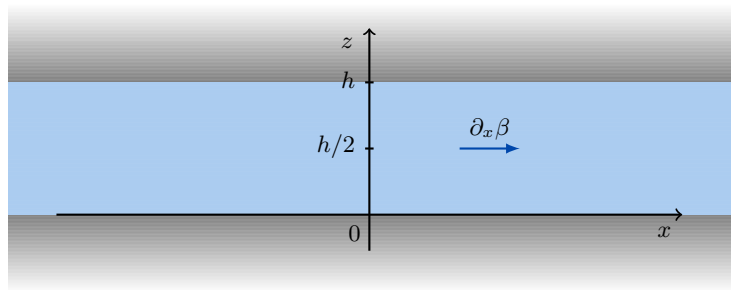

In what follows we will restrict our results to 
the system shown in Fig.~\ref{fig:geoslab}, consisting of a fluid which fills the region between two
infinite parallel planar walls, placed at fixed distance $h$ (the so called {\it slit} or {\it slab} geometry).
Furthermore we impose that the walls behave as hard objects with respect to the fluid and
the fluid is kept out of equilibrium applying a temperature difference in the \hbox{$x$-direction}. 
The temperature difference is set at infinity and is such that the gradient is small and finite.

\subsubsection*{Preliminary assumptions on the symmetry of the solution}
On the basis of the simple geometry of the problem, we {\it expect} that the solutions of the system consisting of 
\hbox{Eq.s~(\ref{eq:stat1})},~(\ref{eq:stat2}) and~(\ref{eq:stat3}) will show some additional 
properties. Here we assume these properties and then we will show that such a solution exist.
The assumptions on the solutions are the following:

\begin{enumerate}
\item The gradient of the field $\beta(\bmr)$ is uniform throughout the fluid 
and is set in the \hbox{$x$-direction}
\begin{equation}
\bm{\nabla}\beta(\bmr)=\big(\partial_x\beta\,,0\,,0\,\big),
\notag
\end{equation}
where $\partial_x\beta$ is a constant. 
\item
The gradient of the field $\mu(\bmr)$ is uniform throughout the fluid                                      
and is set in the \hbox{$x$-direction}
\begin{equation}
\bm{\nabla}\mu(\bmr)=\big(\partial_x\mu\,,0\,,0\,\big).
\notag
\end{equation} 
Within linear response theory the combination of this assumption with the first one 
implies that $\partial_x(\beta\mu)$ is a constant\footnote{The term $\partial_x\beta\partial_x\mu$ is 
of the second order in the gradients of the fields.} and, at linear order in the 
derivatives of the fields, we can write
\begin{equation}
\bm{\nabla}\left[\beta \mu \right](\bmr)=\big(\partial_x(\beta\mu)\,,0\,,0\,\big).
\notag
\end{equation} 
\item
The only \hbox{non-vanishing} component of the velocity field is along the \hbox{$x$-axis}
and is dependent only on the coordinate $z$ normal to the wall
\begin{equation}
\bmu(\bmr)=\big(u^x(z)\,,0\,,0\,\big).
\label{eq:ass3}
\end{equation}
\end{enumerate}
In what follows we apply these assumptions to the conservation laws (\ref{eq:stat1}), (\ref{eq:stat2})
and (\ref{eq:stat3}).

\subsection*{Mass and energy conservation laws}

Due to the symmetries of the system, it turns out that the \hbox{steady-state} conservation law for 
the mass density \hbox{(\ref{eq:stat1})} and for the momentum density (\ref{eq:stat2}) are identically satisfied.

The average value of the momentum density is 
\begin{align}
\big<\hat{j}^{\alpha}(\bmr) \big>&=\left< \hat{\rho}(z)\right>_0u^{x}(z)\,\delta^{\alpha x}+
\int_0^{\infty} \ud t^\prime \int \ud \bmr' \,
\biggl[\big<\hat{j}^{\alpha}(\bmr,t')\, \hat{J}_{\cal H}^x(\bmr')\big>_0 \, \partial_x\beta \notag \\
& \qquad \qquad\qquad \qquad \qquad \qquad \qquad- \beta \big< \hat{j}^{\alpha}(\bmr,t') \,\hat{J}_j^{xz}(\bmr')\big>_0 \, \dep_z u^{x}(z') 
-  \big< \hat{j}^{\alpha}(\bmr,t') \, \hat{j}_{\rho}^{x}(\bmr')\big>_0 \,\partial_{x}\bigl(\beta\mu\bigr)\biggr], 
\notag
\end{align}
where we only made use of the assumptions introduced above.
Due to the symmetry of the {\it equilibrium} system, only the $x$ component of the average momentum density 
is non-vanishing ($\langle j^{y}(\bmr) \rangle=\langle j^{z}(\bmr) \rangle=0$, similar considerations apply to both terms): 
\begin{align}
\big<\hat{j}^{x}(\bmr) \big>&=\left< \hat{\rho}(z)\right>_0u^{x}(z)+
\int_0^{\infty} \ud t^\prime \int \ud \bmr' \,
\biggl[\big< \hat{j}^{x}(\bmr,t')\, \hat{J}_{\cal H}^x(\bmr')\big>_0 \, \partial_x\beta \notag \\
&\qquad \qquad \qquad \qquad \qquad - \beta \big< \hat{j}^{x}(\bmr,t') \,\hat{J}_j^{xz}(\bmr')\big>_0 \, \dep_z u^{x}(z') 
-  \big< \hat{j}^{x}(\bmr,t') \, \hat{j}_{\rho}^{x}(\bmr')\big>_0 \,\partial_{x}\bigl(\beta\mu\bigr)\biggr].
\label{eq:mom_dens_x} 
\end{align}
Therefore the stationarity condition for the mass density (\ref{eq:stat1}) reads 
\begin{equation}
0=\partial_\alpha \big< \hat{j}^{\alpha}(\bmr) \big> 
=\partial_x\int_0^\infty \ud t^\prime \int \ud \bmr' \,
\biggl[\big< \hat{j}^{x}(\bmr,t')\, \hat{J}_{\cal H}^x(\bmr')\big>_0 \, \partial_x\beta 
- \beta \big< \hat{j}^{x}(\bmr,t') \,\hat{J}_j^{xz}(\bmr')\big>_0 \, \dep_z u^{x}(z') 
-  \big< \hat{j}^{x}(\bmr,t') \, \hat{j}_{\rho}^{x}(\bmr')\big>_0 \,\partial_{x}\bigl(\beta\mu\bigr)\biggr].
\notag
\end{equation}
The \hbox{two-point} correlation functions only depend on the difference $x-x'$\footnote{And of course on $z$ and $z'$.} because the averages 
are evaluated at equilibrium and the system is homogeneous along the coordinate $x$. Therefore their integral 
over $\bmr'$ will be independent on $x$ and its derivative vanishes.

Analogous considerations apply to the continuity equation for $\big< \hat{\cal H}(\bmr)\big>$. 
Only the component of the flux along $x$ is different from zero:
\begin{align}
\big< \hat{J}_{\cal H}^{x}(\bmr) \big>&=
\beta \int \ud\bmr^\prime \big<\hat{J}_{\cal H}^{x}(\bmr)\,\hat{j}^{x}(\bmr')\big>_0 \,u^{x}(z')+
\int_0^\infty \ud t^\prime \int \ud \bmr' \,
\biggl[\big< \hat{J}_{\cal H}^{x}(\bmr,t')\, \hat{J}_{\cal H}^x(\bmr')\big>_0 \, \partial_x\beta \notag \\
& \qquad \qquad\qquad \qquad - \beta \big< \hat{J}_{\cal H}^{x}(\bmr,t') \,\hat{J}_j^{xz}(\bmr')\big>_0 \, \dep_z u^{x}(z') 
-  \big< \hat{J}_{\cal H}^{x}(\bmr,t') \, \hat{j}_\rho^{x}(\bmr')\big>_0 \,\partial_{x}\bigl(\beta\mu\bigr)\biggr]. 
\notag
\end{align}
As before, the continuity equation, which in the stationary limit reduces to the derivative w.r.t. $x$ 
of $\big< \hat{J}_{\cal H}^{x}(\bmr) \big>$,
is identically satisfied because the integral of the correlation functions does not depend on $x$.

\subsection*{Momentum conservation law}

The third stationarity condition (\ref{eq:stat3}) gives origin to 
three independent equations. Two of them are identically satisfied whereas the last
one defines the gradient of the velocity profile.
\newline
Let us start with the conservation law for $\big<\hat{j}^{z}(\bmr) \big>$.
The \hbox{$\alpha z$-component} of the momentum current which enters the stationary conservation of momentum density reads
\begin{align}
\big<\hat{J}_j^{\alpha z}(\bmr) \big>=\big< \hat{J}_{j}^{\alpha z}(\bmr)\big>_{LE}+
\int_0^\infty \ud t^\prime \int \ud \bmr' \,
\biggl[\big<\hat{J}_j^{\alpha z}(\bmr,t')\, &\hat{J}_{\cal H}^x(\bmr')\big>_0 \, \partial_x\beta - 
\beta \big< \hat{J}_j^{\alpha z}(\bmr,t') \,\hat{J}_j^{xz}(\bmr')\big>_0 \, \dep_z u^{x}(z')\notag \\
& \qquad \quad -  \big< \hat{J}_j^{\alpha z}(\bmr,t') \, \hat{j}_{\rho}^{x}(\bmr')\big>_0 \,\partial_{x}\bigl(\beta\mu\bigr)\biggr]\notag \\
=\big< \hat{J}_{j}^{\alpha z}(\bmr)\big>_{LE}+
\delta^{\alpha x}\int_0^\infty \ud t^\prime \int \ud \bmr' \,
\biggl[\big<\hat{J}_j^{x z}(\bmr,&t')\, \hat{J}_{\cal H}^x(\bmr')\big>_0 \, \partial_x\beta  - 
\beta \big< \hat{J}_j^{x z}(\bmr,t') \,\hat{J}_j^{xz}(\bmr')\big>_0 \, \dep_z u^{x}(z') \notag \\
&  \quad  \qquad  -  \big< \hat{J}_j^{xz}(\bmr,t') \, \hat{j}_{\rho}^{x}(\bmr')\big>_0 \,\partial_{x}\bigl(\beta\mu\bigr)\biggr],
\label{eq:cons_mom_z_iniziale}
\end{align}
where the last equality follows from the usual symmetry properties. The term proportional to the delta function does not depend on $x$, because the 
correlation functions only depend on $x-x'$, and a suitable change of variable in the integral 
makes it independent on $x$.
\newline
In planar symmetry, the local equilibrium contribution evaluated in Eq.~(\ref{eq:Jj_le_generale}), reduces to
\begin{align}
\big< \hat{J}_{j}^{\alpha z}(\bmr)\big>_{LE}&=
\big<\hat{J}_{j}^{\alpha z}(z)\big>_{0}
-\partial_x \beta\int \ud \bmr'\, x'\,\Bigl[\big<\hat{J}_{j}^{\alpha z}(\bmr)\,\hat{\cal H}(\bmr')\big>_0
-\mu \big<\hat{J}_{j}^{\alpha z}(\bmr)\,\hat{\rho}(\bmr')\big>_0\Bigr] \notag \\
& \qquad \qquad \qquad \qquad \qquad \quad 
+\beta\, \partial_x \mu \int \ud \bmr'\, x'\,\big<\hat{J}_{j}^{\alpha z}(\bmr)\,\hat{\rho}(\bmr')\big>_0
+ C_\mathcal{Q}\big<\hat{J}_{j}^{\alpha z}(z)\big>_0.
\label{eq:Jj_alphazeta_LE_generale}
\end{align}
The quantity $\big< \hat{J}_{j}^{y z}(\bmr)\big>_{LE}$ does not depend on $y$, because on one side the 
equilibrium pressure tensor is diagonal and on the other side the linear corrections, 
which are integrated along $y'$, loose their dependence on $y$. 
\newline
On the other hand $\big< \hat{J}_{j}^{x z}(\bmr)\big>_{LE}$ is different 
from zero due to the presence of the linear contributions in the derivatives of the fields in Eq. (\ref{eq:Jj_alphazeta_LE_generale}).
Furthermore this term depends on $z$, but does not depend on $x$: The integrals in (\ref{eq:Jj_alphazeta_LE_generale}) can be rearranged as
\begin{align}
\int \ud \bmr'\, x'\big<\hat{J}_{j}^{x z}(\bmr)\,\hat{O}(\bmr')\big>_0&=
\int \ud \bmr'\, (x'-x) \big<\hat{J}_{j}^{x z}(\bmr)\,\hat{O}(\bmr')\big>_0
+ x\,\int \ud \bmr'\,\big<\hat{J}_{j}^{x z}(\bmr)\,\hat{O}(\bmr')\big>_0\notag \\
&=\int \ud \bmr'\, (x'-x) \big<\hat{J}_{j}^{x z}(\bmr)\,\hat{O}(\bmr')\big>_0,
\label{eq:elimino_dipdax}
\end{align}
where $\hat{O}(\bmr)$ is one of the scalar operator appearing in~(\ref{eq:Jj_alphazeta_LE_generale}).
Summing up, the \hbox{$xz$~component} of the LE pressure tensor reads
\begin{equation}
\big< \hat{J}_{j}^{x z}(\bmr)\big>_{LE}=
\int \ud \bmr'\,(x-x')\,\Big[ \partial_x \beta \,\big<\hat{J}_{j}^{x z}(\bmr)\,\hat{\cal H}(\bmr')\big>_0
- \partial_x \big(\beta\mu\big) \big<\hat{J}_{j}^{x z}(\bmr)\,\hat{\rho}(\bmr')\big>_0\Big],
\label{eq:Jjxz_LE_planare}
\end{equation}
where we remark that in the linear approach $\partial_x(\beta \mu)=\mu\, \partial_x\beta+\beta\, \partial_x\mu$.
\newline 
Finally, the continuity equation we are considering involves also the LE average of the \hbox{$zz$-component}
of the momentum flux, which can be written as
\begin{equation}
\big< \hat{J}_{j}^{zz}(\bmr)\big>_{LE}=p_{\mathrm{N}}(z)\Big|_{\beta(x),\mu(x)}.
\notag
\end{equation}
Notice that $\big< \hat{J}_{j}^{zz}(\bmr)\big>_{LE}$ depends both on $z$ and $x$, because the equilibrium averages are 
evaluated at the local value of the temperature and chemical potential $\beta(x)$ and $\mu(x)$.
\newline 
Gathering the results obtained so far, the stationarity condition for the $z$ component
of the momentum density reads
\begin{equation}
\partial_\alpha \big< \hat{J}_{j}^{\alpha z}(\bmr)\big>=\partial_z p_{\mathrm{N}}(z)\Big|_{\beta(x),\mu(x)}
=-\frac{1}{m}\rho(z)\Big|_{\beta(x),\mu(x)}\,\partial_z V(z),
\label{eq:eq_idro}
\end{equation}
where $\rho(z)$ is the equilibrium density profile evaluated at the local $\beta(x)$ and $\mu(x)$, whereas for hard walls
$V(\bmr)$ vanishes in the region occupied by the fluid.
Eq.~(\ref{eq:eq_idro}), the so called {\it hydrostatic equilibrium condition}, is always 
fulfilled by the normal component of the pressure tensor at each value of the \hbox{$x$-coordinate} and is not specific to our problem. 
For hard walls~(\ref{eq:eq_idro}) implies that the normal pressure is constant along $z$ and equals the bulk pressure $p$ at the local $\beta(x)$ and $\mu(x)$. 

The stationarity condition for $\big< {j}^{y}(\bmr)\big>$
is identically satisfied because the symmetry of the problem implies $\partial_{\alpha}\big< \hat{J}_{j}^{\alpha y}(\bmr)\big>=0$.

The only non-trivial continuity equation comes from the conservation of the 
\hbox{$x$-component} of the momentum density
\begin{equation}
\partial_\alpha \big< \hat{J}_{j}^{\alpha x}(\bmr)\big>=
\partial_x\big< \hat{J}_{j}^{xx}(\bmr)\big> +\partial_z\big< \hat{J}_{j}^{xz}(\bmr)\big>=0.
\notag
\end{equation}
It is straightforward to show that the relevant terms in this relation can be written as
\begin{align}
\big<\hat{J}_j^{\alpha x}(\bmr) \big>&=\big< \hat{J}_{j}^{\alpha x}(\bmr)\big>_{LE}+
\delta^{\alpha z}\int_0^\infty \ud t^\prime \int \ud \bmr' \,
\biggl[\big<\hat{J}_j^{x z}(\bmr,t')\, \hat{J}_{\cal H}^x(\bmr')\big>_0 \, \partial_x\beta 
- \beta \big< \hat{J}_j^{x z}(\bmr,t') \,\hat{J}_j^{xz}(\bmr')\big>_0 \, \dep_z u^{x}(z') \notag \\
&\qquad \qquad \qquad \qquad \quad \qquad \qquad \qquad \qquad \qquad \qquad \qquad \quad \qquad 
-  \big< \hat{J}_j^{x}(\bmr,t') \, \hat{j}_{\rho}^{x}(\bmr')\big>_0 \,\partial_{x}\bigl(\beta\mu\bigr)\biggr], 
\notag
\end{align}
where the tangential pressure acquires a dependence on $x$ and reads 
\begin{equation}
\big< \hat{J}_{j}^{xx}(\bmr)\big>_{LE}=p_\mathrm{T}(z)\Big|_{\beta(x),\mu(x)}
\notag
\end{equation}
and $\big< \hat{J}_{j}^{zx}(\bmr)\big>_{LE}$ has already been evaluated in Eq.~(\ref{eq:Jjxz_LE_planare}).
\newline
Finally, the stationarity condition for the \hbox{$x$-component} of the momentum density reads
\begin{align}
0=\partial_x p_\mathrm{T}(z)\Big|_{\beta(x),\mu(x)}&+\partial_z\int \ud \bmr'\,(x-x')\,\bigg[ \partial_x \beta \,\big<\hat{J}_{j}^{x z}(\bmr)\,\hat{\cal H}(\bmr')\big>_0
- \partial_x \big(\beta\mu\big) \big<\hat{J}_{j}^{x z}(\bmr)\,\hat{\rho}(\bmr')\big>_0\bigg]\notag\\
&+\partial_z\int_0^\infty \ud t^\prime \int \ud \bmr' \,
\biggl[\big<\hat{J}_j^{x z}(\bmr,t')\, \hat{J}_{\cal H}^x(\bmr')\big>_0 \, \partial_x\beta 
-\big< \hat{J}_j^{xz}(\bmr,t') \, \hat{j}_{\rho}^{x}(\bmr')\big>_0 \,\partial_{x}\bigl(\beta\mu\bigr) \notag \\ 
&\qquad \qquad \qquad \qquad \qquad\qquad\qquad- \beta \big< \hat{J}_j^{x z}(\bmr,t') \,\hat{J}_j^{xz}(\bmr')\big>_0 \, \dep_z u^{x}(z') \biggr]. 
\label{eq:cons_mom_x_fin}
\end{align}
This condition is an integral equation for the velocity profile $u^{x}(z)$. 

\subsection{Open channel}

The continuity equation for the momentum density along the \hbox{$x$-direction}~(\ref{eq:cons_mom_x_fin}) 
can be integrated provided we restrict to the free flow in a {\it infinitely long channel} (see Fig.~\ref{fig:geoslab}),
large enough to guarantee that in the central region the perturbation due to the walls is absent.
In this region the fluid can be considered homogeneous and isotropic: The normal and the tangential 
components of the pressure tensor coincide and reduce to the bulk pressure $p$ of the fluid evaluated at the 
given value of the fields $\beta(x)$ and $\mu(x)$
\begin{equation}
p_{\mathrm{N}}(z\sim h/2)\Big|_{\beta(x),\mu(x)}=p_{\mathrm{T}}(z\sim h/2)\Big|_{\beta(x),\mu(x)}=p\,\big|_{\beta(x),\mu(x)}.
\notag
\end{equation}
We remark again that when a temperature gradient is applied to the system the bulk pressure becomes \hbox{$x$-dependent}. 

The free flow in an {\it open} channel imposes equal bulk pressure at the left
and the right boundaries. It follows that the temperature and the chemical potential
gradients will adapt in order to guarantee that 
\begin{equation}
0=\partial_x p\,\big|_{\beta(x),\mu(x)} = 
\partial_x\beta\,\left [ \partial_\beta p + \frac{\partial_x\mu}{ \partial_x\beta}\,\partial_\mu p\right ].
\notag
\end{equation}
This equation fixes the ratio between the temperature and the chemical potential gradients, which can 
be expressed in terms of thermodynamic densities as
\begin{equation}
\frac{\partial_x \mu}{\partial_x \beta}= -\frac{\partial_\beta p }{\partial_\mu p}=k_\mathrm{B}\,T^2\,\frac{s}{\rho},
\notag
\end{equation}
where $s$ is the entropy density and we made use of the thermodynamic relations\footnote{Here we apply without ambiguity  the formalism specific of macroscopic 
thermodynamics because in the central region the system is supposed to be homogeneous.}
\begin{equation}
\partial_T p\,\big|_{\mu}=-\partial_T (\Omega/V)\big|_{\mu}=s, \qquad \qquad \qquad \partial_\mu p\,\big|_{T}=-\partial_\mu (\Omega/V)\big|_{T}=\rho.
\notag
\end{equation}
Here the derivatives of the grand potential $\Omega$ are always evaluated at constant volume $V$.
The condition obtained above allows to express $\partial_x(\beta\mu)$ more naturally as
\begin{equation}
\partial_x(\beta\mu)=\partial_x\beta \left(\mu+T\,\frac{s}{\rho} \right)=\partial_x\beta \frac{u+p}{\rho}=\partial_x\beta\,h_m,
\label{eq:bc_derbetamu}
\end{equation}
where $u$ is the internal energy and $h_m$ is the bulk enthalpy per unit mass.

Now the stationarity condition~(\ref{eq:cons_mom_x_fin}) can be straightforwardly integrated 
along $z$ from $0$ to $h/2$, where $h$ is the distance between the walls.
All the integrals derived w.r.t. $z$ in~(\ref{eq:cons_mom_x_fin}) are odd functions of 
$z$ with respect to $h/2$ and, if evaluated at this point, vanish.
Therefore the \hbox{integro-differential} equation for the velocity profile can be written as Eq. (7) of the main text, that we report here for future 
reference:
\begin{equation}
\int_0^h\ud z^\prime {\cal K}(z,z^\prime) \, \dep_{z}\, u^x(z^\prime) =
\partial_x\beta \, {\cal S}(z),
\label{eq:equ}
\end{equation}
where the kernel ${\cal K}(z,z^\prime)$ reads
\begin{equation}
{\cal K}(z,z^\prime) = \beta\,\int_0^\infty \ud t'\int\ud\bmr_\perp^\prime 
\big <\hat J_j^{xz}(\bmr,t')\, \hat J_j^{xz}(\bmr^\prime)\big >_0
\nonumber 
\end{equation}
and the source term ${\cal S}(z)$ can be expressed as the sum of a static and dynamic contributions ${\cal S}(z)={\cal S}_s(z)+{\cal S}_d(z)$
\begin{align}
{\cal S}_s(z) &=\int^z_{h/2} \ud z^\prime \left.\frac{\partial p_{\mathrm{T}}(z^\prime)}{\partial \beta}\right|_p  -
\int \ud\bmr^\prime \, (x-x^\prime)\,\big< \hat J_j^{xz}(\bmr) \, \hat{\cal P}(\bmr^\prime) \big>_0, 	\label{eq:static} \\
{\cal S}_d(z) &= \int_0^\infty \ud t'\int\ud\bmr^\prime 
\big<\hat J_j^{xz}(\bmr,t')\, \hat J_Q^{x}(\bmr^\prime)\big >_0. \qquad
\label{eq:dynamic}
\end{align}
Here we defined the operator $\hat{\cal P}(\bmr)=h_m\,\hat\rho(\bmr) - \hat{\cal H}(\bmr)$ and the 
operator $\hat J_Q^{\alpha}(\bmr) = \hat J_{\cal H}^{\alpha}(\bmr)-h_m\,\hat j_{\rho}^{\alpha}(\bmr)$ 
which can be interpreted as a mechanism of heat flux\footnote{Notice that, 
strictly speaking, this definition of the heat flux does not correspond to the microscopic counterpart of the heat flux introduced within 
classical hydrodynamics. Irving and Kirkwood have shown~\cite{irvingkirkwood_1949} that, from a microscopic point of view,  
it is possible to introduce the mean velocity $\bmv(\bmr)$ 
$$
\big<\hat{\rho}(\bmr,t)\big>\bmv(\bmr,t)=\big<\hat{\bmj}(\bmr,t)\big>
$$
and then the heat flux is defined through the average energy conservation 
$$ 
\partial_t \big<\hat{\cal H}(\bmr,t)\big> =-\partial_\alpha \hat J_{\cal H}^{\alpha}(\bmr,t)= - \partial_\alpha \left[v^{\alpha}(\bmr,t)\big<\hat {\cal H}(\bmr,t)\big> +
\big<\hat J_{\cal Q}^{\alpha}(\bmr,t)\big> + v^{\beta}(\bmr,t)\big<\hat J_j^{\alpha \beta}(\bmr,t)\big>\right].
$$}.
Notice that the static contribution has been rewritten regarding the tangential pressure as a function of 
the $\beta(x)$ and of the bulk pressure $p$. Indeed, at midpoint 
\begin{equation}
p_\mathrm{T}(h/2)\Big|_{\beta(x),\mu(x)}\sim p
\notag
\end{equation}
for each value of the coordinate $x$ and we can replace the local chemical potential with the bulk pressure obtaining
\begin{equation}
\partial_x p_\mathrm{T}(z)\big|_{\beta(x),p}=\left. \frac{\partial p_\mathrm{T}(z)}{\partial \beta}\right|_p \partial_x \beta.
\label{eq:gradp_gradt_constpress}
\end{equation}

The solution of this set of equations provides the {\it gradient} $\partial_z\,u^x(z)$ of the velocity field $u^x(z)$,
which does not have a direct physical meaning. The real flow is related to the average value of the
mass current, Eq. (\ref{eq:mom_dens_x}), which, for an open channel, reads
\begin{equation}
\big<\hat j^x(z)\big> = \rho(z)\,u^x(z) + \int_0^\infty \ud t' \int \ud\bmr^\prime\,
\Bigr[ \big< \hat j^x(\bmr,t') \,\hat J_Q^x(\bmr^\prime)\big>_0 \,\partial_x\beta  
- \beta\,\big< \hat j^x(\bmr,t')\, \hat J_j^{xz}(\bmr^\prime)\big>_0\, \dep_{z} u^x(z')
\Bigl].
\label{eq:momenth2}
\end{equation}
In this expression for the mass current both the velocity field and its derivative appear. The actual flow 
can be determined only by imposing a physical boundary condition to the average mass current in some point (see the discussion in the main text).

\section{Thermo-osmosis in liquids}

In this Section we evaluate Eq.s~(\ref{eq:equ}) and~(\ref{eq:momenth2}),
exact at least for sufficiently small perturbations from equilibrium, under the approximations 
commonly accepted in liquids (see e.g.~\cite{surface_forces_1987}). The purpose is to check if our result for the the slip velocity 
in an open channel (see Fig.~\ref{fig:geoslab})
\begin{equation}
v^x(z)=\frac{\big<\hat j^x(z)\big>}{\rho(z)}
\notag
\end{equation}
reduces, far from the walls, to the well known expression 
\begin{equation}
v_{\infty}=\frac{1}{\eta}\int_{0}^{\infty}\ud z \, z\,\Delta h(z) \,\frac{\nabla T}{T},
\label{eq:derjaguin}
\end{equation} 
where $\eta$ is the bulk viscosity and $\Delta h(z)= h(z)-h_b$ is the excess (with respect to bulk) local enthalpy at a given height $z$.
This result was derived for the first time by Derjaguin and Sidorenko in 1941~\cite{derjaguin_sidorenkov_1941thermoosmosis}
on the basis of nonequilibrium thermodynamics. 
\newline 
An expression {\it similar} to~(\ref{eq:derjaguin}) has been obtained in a recent work 
by Ganti et al. \cite{ganti_2017}, where the enthalpy difference is defined as $\Delta h(z)=\rho(z) (h_m(z)-h_m)$, 
with $h_m(z)$ the enthalpy per unit mass of the fluid at distance $z$ from the wall.
In this work the \hbox{thermo-osmotic} slip has been evaluated by applying the \hbox{Gibbs-Duhem} relation, 
valid for homogeneous systems at equilibrium, to each fluid layer at height $z$. Here the underlying assumption is that
a local density approximation can be applied to each stratification at height $z$, where, according to the authors,
the system can be considered homogeneous with constant density equal to the local value $\rho(z)$. 
A similar argument has been adopted also in Ref.~\cite{surface_forces_1987}.

In the spirit of these continuum approaches we evaluate all the correlation functions in~(\ref{eq:equ}) and~(\ref{eq:momenth2}) in the bulk 
and we assume that the kernel is a short-ranged function
\begin{equation}
{\cal K}(z,z^\prime) \sim \eta\,\delta(z-z^\prime),
\label{eq:kernel_bulkviscosity}
\end{equation}
with $\eta$ the bulk viscosity.
The first assumption also imply that the dynamic source term~(\ref{eq:dynamic}) vanishes, because in homogeneous systems tensors  
preserving isotropy and homogeneity must be proportional to the identity.
For the same reason the static source~(\ref{eq:static}) retains only the contribution
including the anisotropy of the pressure tensor. Summing up, the source term within this ``Derjaguin'' approximation reads 
\begin{equation}
{\cal S}_{\mathrm{Derj}}(z)=\int^z_{h/2} \ud z^\prime \left.\frac{\partial p_\mathrm{T}(z^\prime)}{\partial \beta}\right|_p 
\notag
\end{equation}
and the differential equation for the velocity profile is
\begin{equation}
\partial_z u^x(z) =-\frac{\partial_x\beta}{\eta} \int_z^{h/2} \ud z^\prime \left.\frac{\partial p_\mathrm{T}(z^\prime)}{\partial \beta}\right|_p .
\label{eq:equliquid}
\end{equation}
On the other hand, the momentum density~(\ref{eq:momenth2}), which is the physical quantity related to the real mass flux, 
only retains the linear contribution in the velocity profile $u^x$
\begin{equation}
\big<\hat j^x(z)\big> = \rho(z)\,u^x(z).
\label{eq:jliquids}
\end{equation}
The integration of the \hbox{first-order} differential equation~(\ref{eq:equliquid}) needs a boundary condition (see the discussion in the main text).
Here we adopt the \hbox{no-slip} boundary condition $\big<\hat j^x(0)\big> = 0$,
which implies $u^x(0)=0$. Once this choice is made, the mass flux reads\footnote{After the straightforward change of variable
\begin{equation}
\int_0^{z} \ud x \int_x^{h/2} \ud y \,f(y) = \int_0^{h/2}\ud y \, f(y) \int_0^{\mathrm{Min}(y,z)}\ud x.
\notag
\end{equation}} \begin{equation}
\big<\hat j^x(z)\big> =-\frac{\rho(z)\partial_x\beta}{\eta} \int_0^{h/2} \ud z^\prime \, 
\mathrm{Min}(z,z')\left.\frac{\partial p_\mathrm{T}(z^\prime)}{\partial \beta}\right|_p.
\notag
\end{equation}
In the asymptotic limit, i.e. when $h$ and $z$ are larger than the typical length scale 
of the correlations ($z\to \infty$ and $h\to \infty$), $\mathrm{Min}(z,z')\sim z'$ and the slip velocity can be written as
\begin{equation}
v_{\infty}=\frac{\big<\hat j^x\big>_{\infty}}{\rho_b} =-\frac{\partial_xT}{\eta}\left.\frac{\partial }{\partial T}\right|_p 
\int_0^{\infty} \ud z^\prime \, z \, \big[p_\mathrm{T}(z^\prime)-p\big]=-\frac{\partial_xT}{\eta}\left.\frac{\partial }{\partial T}\right|_p 
\int_0^{\infty} \ud z^\prime \, z \, \Delta p_\mathrm{T}(z^\prime).
\label{eq:nostroderjaguin}
\end{equation} 
In this expression the bulk pressure in the integral has been subtracted and does not provide an additional contribution 
because the derivative is taken at fixed  bulk pressure $p$.

\subsubsection*{Back to Derjaguin's (and Ganti's) result}

The expression for the asymptotic slip velocity~(\ref{eq:nostroderjaguin}), valid under the hypotheses introduced above, can be related 
to Derjaguin's (and Ganti's~\cite{ganti_2017}) prediction~(\ref{eq:derjaguin}). 
Indeed, in the spirit of the local density approximation, underlying both Ref.s~\cite{surface_forces_1987} and~\cite{ganti_2017}, we
assume that in a liquid layer at distance $z$ from the wall the tangential pressure is equal to the pressure of a homogeneous system where the 
density equals the local density $\rho(z)$ and the temperature and the chemical potential are fixed at the values of $\beta(x)$ and $\mu(x)$ 
respectively\footnote{In this way the tensorial character of the pressure in inhomogeneous regions is completely lost, 
and the pressure turns out to be uniquely defined.}.
Under this hypothesis, the pressure gradient in the $x$-direction in Eq.~(\ref{eq:nostroderjaguin})\footnote{Notice that, 
according to Eq.~(\ref{eq:gradp_gradt_constpress}), 
$$
\left.\frac{\partial \big[p_\mathrm{T}(z)-p\big] }{\partial T}\right|_p \partial_x T= 
\left.\frac{\partial \big[p_\mathrm{T}(z)-p\big] }{\partial \beta}\right|_p \partial_x \beta
=\partial_x p_\mathrm{T}(z)\big|_{\beta(x),p}=\partial_x p_\mathrm{T}(z)\Big|_{\beta(x),\mu(x)}=\partial_x p_\mathrm{T}(z)\Big|_{T(x),[\beta\mu](x)},$$
and the last identities follow from the freedom in the choice of the thermodynamic variables.} reads
\begin{align}
\partial_x p_\mathrm{T}(z)\Big|_{T(x),[\beta\mu](x)}&=\left.\frac{\partial p(z)}{\partial T}\right|_{[\beta\mu](x)} \partial_x T+
\left.\frac{\partial p(z)]}{\partial \beta\mu}\right|_{T(x)} \partial_x [\beta\mu]\notag\\
&=\frac{h(z)}{T}\partial_x T +\frac{\rho(z)}{\beta}\partial_x [\beta\mu]\notag \\
&=\frac{h(z)- h_m\, \rho(z) }{T} \, \partial_x T \notag
\end{align}
where, according to the local density approximation, $p(z)$ and $h(z)$ are the pressure and the enthalpy per unit volume respectively 
of a {\it homogeneous} system at density $\rho(z)$\footnote{Notice that the dependence of the thermodynamic variables on the local inverse temperature $\beta(x)$ 
and chemical potential $\mu(x)$ is understood.} 
and the last equality follows from Eq.~(\ref{eq:bc_derbetamu}). 

\subsubsection*{Connection with the Navier-Stokes equations}

It is possible to show that~(\ref{eq:nostroderjaguin}) coincides with the solution of the linearised Navier-Stokes 
equation for an incompressible fluid when a tangential pressure gradient given by the LE expression is applied 
\cite{ganti_2017,piazzaparola_2008}. The fully macroscopic Navier-Stokes approach based on the continuum approximation 
states that the differential equation obeyed by the stationary velocity field ${\bm v(\bmr)}$ of an incompressible 
fluid can be written as 
\begin{equation}
0=\partial_t (\rho v^\alpha)=-\partial_\beta \Pi^{\alpha \beta} + F^{\alpha}.
\notag
\end{equation}
This equation is the (stationary) Navier-Stokes equation, where $\bm{F}$ is the force field acting on the fluid,
which may be due either to the presence of the wall or to an external field.
The momentum flux tensor $\Pi^{\alpha \beta}$ can be written in terms of the stationary 
momentum flux $\big<\hat{J}_j^{\alpha \beta}(\bmr) \big>=\pi^{\alpha\beta}$ introduced above as
\begin{equation}
\Pi^{\alpha \beta}=\pi^{\alpha\beta}+\rho v^\alpha v^\beta-\eta\left[\frac{\partial v^\alpha}{\partial r^\beta}-\frac{\partial v^\beta}{\partial r^\alpha}\right].
\notag
\end{equation}
In the limit of small velocities and within the adopted planar symmetry, the
the \hbox{$x$-component} of the velocity field fulfils the linearised \hbox{Navier-Stokes} equation
\begin{equation}
\eta \frac{\ud^2 v^x}{\ud z^2}=\partial_x p_\mathrm{T},
\notag
\end{equation}
where the \hbox{wall-fluid} interaction is modelled as a hard-core potential (${\bm F}$ is vanishing in the fluid domain). 
Following the same line of reasoning as before, the last contribution can be written in terms of the temperature derivative of
the tangential pressure at fixed bulk pressure:
\begin{equation}
\eta \frac{\ud^2 v^x}{\ud z^2}=\partial_xT \left.\frac{\partial p_\mathrm{T}(z)}{\partial T}\right|_p .
\label{eq:linearized_ns}
\end{equation}
Imposing \hbox{no-slip} boundary conditions at the wall ($v^x(0)=0$) Eq.~(\ref{eq:linearized_ns}) can be easily 
integrated and the asymptotic velocity field is\footnote{The first integration
is from $h/2$ to $0$, and we exploit the symmetry of the problem which implies that the 
derivative of the velocity profile vanishes in the middle point between the walls. Then, the second integration
proceeds as already described.} 
\begin{equation}
v_{\infty}= -\frac{\partial_xT}{\eta}\left.\frac{\partial }{\partial T}\right|_p 
\int_0^{\infty} \ud z^\prime \, z \, \big[p_\mathrm{T}(z^\prime)-p\big],
\notag
\end{equation}
which coincides with Eq.~(\ref{eq:nostroderjaguin}).

\section{Thermo-osmosis in gases}

In the {\it ideal gas} limit, i.e. ignoring the interparticle interactions, the momentum~(\ref{eq:mom_flow}) and 
the energy~(\ref{eq:form_def_JH}) fluxes reduce to:
\begin{equation}
\hat{J}_j^{\alpha\beta}(\bmr)=\sum_i \frac{p_i^{\alpha}p_i^{\beta}}{m} \delta(\bmq_i-\bmr), \qquad \qquad 
\hat{J}_{\cal H}^\alpha(\bmr)=\sum_i\frac{p_i^2\,p^{\alpha}_i}{2m^2}\delta(\bmq_i-\bmr).
\notag
\end{equation}
In this Section we provide some details of the evaluation of the mass current~(\ref{eq:mom_dens_x}) induced by the thermal gradient. 
We begin by solving the \hbox{integro-differential} equation for the velocity profile $u^x(z)$~(\ref{eq:equ}). 
Then we obtain the \hbox{thermo-osmotic} velocity by evaluating the additional dynamic terms appearing in the mass current~(\ref{eq:mom_dens_x}).

\subsection*{Solution of the equation for $u^x(z)$}

The static source term ${\cal S}_s(z)$ vanishes because for ideal gases $p_{\mathrm{T}}=p_{\mathrm{N}}=p$
and the equilibrium average in~(\ref{eq:static}) is performed on a quantity which is odd in the momenta.
Then, the source term reduces to
\begin{equation}
{\cal S}(z)= 
\sum_{i,l}  \left < 
\int_0^\infty dt\int\ud \bmr' \, \delta\big(\bmr-\bmq_l(t)\big)\, \delta\big(\bmr'-\bmq_i\big)
\frac{p_l^x(t)\,p_l^z(t)}{m^2} 
\, p_i^x \left[
\frac{p_i^2}{2m} - m h_m \right] \right >_0,
\label{eq:idealgas_source0}
\end{equation}
where the equilibrium average is evaluated according to the equilibrium distribution~(\ref{eq:fequnder})
and $\bmq_i$ and $\bmp_i$ are the coordinate and the momentum of the particle at $t=0$ respectively. 
\newline
Without any kind of interaction between the particles the time integral in~(\ref{eq:idealgas_source0}) is diverging 
because the correlations persist at all times. In order to mimic the behaviour of an almost ideal gas, where some collisions appear, 
we introduce a finite relaxation time $\tau$. This procedure introduces the collisions between the (ideal) particles {\it a posteriori}, 
and $\tau$ is by definition the time interval between two collisions of a given particle.
In addition, only the contribution arising from the same particle (i.e. $i=l$) is \hbox{non-vanishing} and the source term reads
\begin{equation}
{\cal S}(z)= 
\sum_i \bigg< 
\int_0^\tau dt\int\ud \bmr' \, \delta\big(\bmr-\bmq_i(t)\big)\, \delta\big(\bmr'-\bmq_i\big)
\frac{p_i^x(t)\, p_i^z(t)}{m^2}  \, p_i^x
\left[\frac{p_i^2}{2m} - m h_m \right] \bigg>_0.
\label{eq:idealgas_source1}
\end{equation}
In the case of a perfectly reflecting wall, it is straightforward to show that the source term is zero. 
Indeed, specular reflections without energy exchange conserve both the \hbox{$x$-component} and the modulus of the momentum.
It follows that all the integrated quantities in Eq.~(\ref{eq:idealgas_source1}) can be evaluated at time $t$.
If we perform the canonical transformation ${\scr U}(-t)$ the average over the momenta does not depend on time, and 
the source term vanishes.

In agreement with the results obtained within kinetic theory \cite{kennard_1938kinetic}, the occurrence of
thermal creep is possible only assuming that in the \hbox{particle-surface} scattering the momentum or the energy 
are not conserved\footnote{Notice that ${\cal S}(z)=0$ implies $\partial_z u^x(z)=0$, which means $u^x(z)={\rm const}$. The actual 
value of this constant is determined by the boundary condition.}.
\newline
In order to mimic this behaviour and to obtain analytical results we assume that, due to the interaction with the surface 
during the scattering process, the \hbox{$x$-component} of the particle's momenta before and after the collision are fully uncorrelated.
Furthermore, we restrict to the \hbox{semi-infinite} geometry, where only the wall at $h=0$ (see Fig.~\ref{fig:geoslab}) is present,  
in order to avoid multiple collisions between the surfaces. 
\newline 
The averages can be evaluated without any loss in generality within the canonical ($N,V,T$) ensemble and the source term reads
\begin{equation}
{\cal S}(z)= 
\frac{N}{\tilde{\cal Q}_{\mathrm{c}}}\int_0^\tau dt\int\ud \bmr' \int\ud \bmq \int\ud \bmp \, \, \delta\big(\bmr-\bmq(t)\big)\, \delta\big(\bmr'-\bmq\big)
\frac{p^x(t)\,p^z(t)}{m^2}  \, p^x
\left[\frac{p^2}{2m} - m h_m \right] e^{-\beta \nicefrac{p^2}{2m}},
\label{eq:idealgas_source2}
\end{equation}
where $\tilde{\cal Q}_{\mathrm{c}}=V(2\pi m k_{\mathrm{B}}T)^{\nicefrac{3}{2}}$ and the factor $N$ takes into account that 
the contributions in~(\ref{eq:idealgas_source1}) arising from different particles are equal.
\newline
In order to evaluate the source term, let us briefly examine the behaviour of a particle   
before a given time $t$ and for a set of initial coordinates $\bmq$ and $\bmp$. 
If $p^z\ge -\,m {q^z}/{t}$, the particle does not bounce on the wall in the time interval $[0,t]$ and we can write
\begin{equation}
\bmp(t)=\bmp; \qquad \qquad  \qquad \qquad  \bmq(t)=\bmq+\frac{\bmp}{m}t.
\label{eq:time_scattering}
\end{equation}
On the other hand, when $p^z < -\,m{q^z}/{t}$ the particle hits the wall at time $t_s<t$.  
During the scattering the particle has completely lost the memory of the value of $p^x$ before the 
bounce, therefore its self correlation is equal to $0$ and the contribution in~(\ref{eq:idealgas_source2})  
arising from $p^z < -\,m{q^z}/{t}$ vanishes.
Therefore we can restrict the integral over $p^z$ to the set $[-\,m \,{q^z}/\,{t},+\infty]$ and, according to Eq.~(\ref{eq:time_scattering}) 
we can write
\begin{equation}
{\cal S}(z)= 
\frac{N}{\tilde{\cal Q}_{\mathrm{c}}}\int_0^\tau dt \int\ud \bmr' \int\ud \bmq \int \ud \bmp_{\perp} 
\int_{-mq^z/{t}}^{+\infty} \ud p^z \, \delta\biggl(\bmr-\bmq-\frac{\bmp}{m}t\biggr)\, \delta\big(\bmr'-\bmq\big)
\frac{\big(p^x\big)^2\,p^z}{m^2} \left[\frac{p^2}{2m} - m h_m \right] e^{-\beta \nicefrac{p^2}{2m}},
\notag
\end{equation}
where the integral over the momentum $\bmp_{\perp}$ orthogonal to $p^z$ is extended to $\mathbb{R}^2$.
The final result for the source term, after a careful evaluation of the remaining integrals, reads
\begin{equation}
{\cal S}(z)= - \frac{N  \pi m \tau}{\tilde{\cal Q}_{\mathrm{c}}\,\beta^4}\exp\Biggl[-\beta \,\frac{m z^2}{2\tau^2}\Biggr].
\notag
\end{equation}
Similar arguments allow to express the kernel as
\begin{equation}
{\cal K}(z,z')= \frac{N 2 \pi m^2}{\tilde{\cal Q}_{\mathrm{c}}\, \beta^2} \Theta(z)\Theta(z')\,\exp\left[-\beta \,\frac{m \,\left(z-z'\right)^2}{2\tau^2}\right],
\notag
\end{equation}
where $ \Theta(\cdot)$ is the Heaviside function.
Performing an appropriate change of variables, the differential equation~(\ref{eq:equ}) for $u^x(z)$ can be written as
\begin{equation}
\int_{0}^{+\infty}\ud z' \, \dep_z u^x(z') \frac{2m}{\tau k_{\mathrm{B}} \partial_x T} e^{-z'^2+2\zeta z'}=1,
\notag
\end{equation}
where $\zeta=z\sqrt{m\beta/2\tau^2}$. The solution can be determined up to an additive constant $\mathrm{C}$
and reads
\begin{equation}
u^x(z)=\frac{k_{\mathrm{B}}}{2m} \,\tau\, \Theta(z+\delta)\, \partial_x T + \mathrm{C},
\notag
\end{equation}
where $\delta\to0^+$ and the constant can be fixed imposing the boundary condition for the mass current.
The relaxation time introduced above can be related to the {\it bulk} viscosity $\eta$, which appears in 
most of the expressions for the \hbox{thermo-osmotic} flow present in the literature~\cite{maxwell_1879,kennard_1938kinetic}, 
and can be defined in terms of $\tau$ as
\begin{equation}
\eta=\beta \int_0^\tau \ud t\int\ud \bmr' \, \big <\hat J_j^{xz}(\bmr,t)\, \hat J_j^{xz}(\bmr^\prime)\big >_0.
\label{eq:visco_gas}
\end{equation}
The integrals in~(\ref{eq:visco_gas}) can be evaluated making use of the same arguments introduced above, and, after 
simple algebra, we obtain $\eta= p \,\tau $.
Finally, far from the wall the field $u^x(z)$ can be written as 
\begin{equation}
u^x(z)=\frac{\eta}{p}\frac{k_{\mathrm{B}} T}{2m}\frac{\partial_x T}{T}+\mathrm{C}=\frac{\eta}{2\rho}\frac{\partial_x T}{T}+ \mathrm{C}.
\label{eq:finalu_const}
\end{equation}

\subsection*{Mass current}
As already stated, the velocity field does not have a direct physical meaning: The real flow is related 
to the average value of the $x$ component of the mass current (Eq.~(\ref{eq:momenth2}) and Eq.~(10) of the main text), which we report here:
\begin{equation}
\big<\hat j^x(z)\big> = \rho\,u^x(z) + \int_0^\tau \ud t \int \ud\bmr^\prime\,
\bigg\{\Big< \hat j^x(\bmr,t) \,\Big[\hat J_{\cal H}^x(\bmr^\prime)- h_m \hat{j}^x(\bmr') \Big]\Big>_0 \,\partial_x\beta 
- \beta\,\big< \hat j^x(\bmr,t)\, \hat J_j^{xz}(\bmr^\prime)\big>_0\, \dep_{z} u^x(z')
\bigg\},
\label{eq:momenth2e}
\end{equation}
where $\rho(z)$ equals the bulk density for an ideal fluid and the flux $\hat J_Q^x$ has been written explicitly.
It is straightforward to prove that in the case of perfectly reflecting hard walls the mass current vanishes. As shown before, the velocity profile 
is equal to zero and for this reason the first and the last contributions in~(\ref{eq:momenth2e}) vanish, whereas the remaining terms
exactly cancel. Therefore, in order to obtain a net \hbox{thermo-osmotic} flow can must impose, as done before, a scattering process inducing an exchange of 
momentum between the particle and the wall. In doing so, the first contribution in~(\ref{eq:momenth2e}) is trivial, 
whereas, after some algebra, the integrals over $\bmr'$ of the dynamic correlation functions read
\begin{eqnarray}
\int \ud\bmr^\prime\,\big< \hat j^x(\bmr,t)\hat J_{\cal H}^x(\bmr^\prime) \big>_0&=&
\frac{5}{2} \frac{N\pi m^2}{\tilde{\cal Q}_{\mathrm{c}}\beta^3} \frac{2\pi}{m\beta}\left[\mathrm{erf}\left(z\sqrt{\frac{\beta m}{2t^2}}\right) 
+1\right] - \frac{N\pi m^2}{\tilde{\cal Q}_{\mathrm{c}}\beta^3} \frac{z}{t} \exp\Biggl(-\beta \,\frac{m z^2}{2t^2}\Biggr),\notag \\
h_m \int \ud\bmr^\prime\,\big< \hat j^x(\bmr,t)  \hat{j}^x(\bmr') \big>_0&=&
\frac{5}{2} \frac{N\pi m^2}{\tilde{\cal Q}_{\mathrm{c}}\beta^3} \frac{2\pi}{m\beta}\left[\mathrm{erf}\left(z\sqrt{\frac{\beta m}{2t^2}}\right) 
+1\right],          \notag \\
\int \ud\bmr^\prime\,\big< \hat j^x(\bmr,t)\, \hat J_j^{xz}(\bmr^\prime)\big>_0\, \dep_{z} u^x(z')&=&
\frac{N m^2 \pi k_{\mathrm{B}}\, \tau \partial_xT }{\tilde{\cal Q}_{\mathrm{c}}\beta^2 } \,\frac{z}{t^2}\, \exp\Biggl(-\beta \,\frac{m z^2}{2t^2}\Biggr).
\notag
\end{eqnarray}
The final result for the mass current, after the time integration is, for $z>0$,
\begin{equation}
\big<\hat j^x(z)\big> = \frac{\eta}{2}\frac{\partial_x T}{T} + \frac{\eta}{4}
\left\{\mathrm{erf} \left(\sqrt{\frac{3}{2}}\frac{z}{\ell_g}\right)-
\sqrt{\frac{3}{2\pi}}\,\frac{z}{\ell_g}\, \mathrm{Ei}\left[-\frac{3}{2}\left(\frac{z}{{\ell}_g}\right)^2\right]\right\} 
\notag
\end{equation}
where we have imposed \hbox{no-slip} boundary conditions for the mass current at $z=0$\footnote{The constant in~(\ref{eq:finalu_const}) 
reads ${\rm C}=\frac{\eta}{4\rho}\frac{\partial_x T}{T}$.}, 
$\mathrm{Ei}(\cdot)$ is the exponential integral and $\ell_g\,=\,\tau \sqrt{3/(m\beta)}$.
Far from the walls ($z\gg\ell_g$) the exponential integral rapidly decays to $0$ and the slip velocity $v_{\infty}=\big<\hat j^x(z)\big>|_{z\gg\ell_g}/\rho$ reduces to
\begin{equation}
v_{\infty} =\frac{3}{4}\,\frac{\eta}{\rho}\,\frac{\partial_x T}{T}=\frac{3}{4} \,k_{\mathrm{B}}T \, \frac{\eta}{p}\,\frac{\partial_x T}{T}.
\notag
\end{equation}

\bibliographystyle{apsrev4-1}

\bibliography{bib}